\PassOptionsToPackage{table}{xcolor}
\documentclass[sigconf,screen]{acmart}

\newif\ifpreprint
\preprinttrue   

\usepackage{tabularx}
\usepackage{booktabs}
\usepackage{dcolumn} 
\newcolumntype{d}[1]{D{.}{.}{#1}}

\usepackage{subcaption}
\usepackage{enumitem}
\setlist[enumerate,1]{label=\textbf{RQ\arabic*}, leftmargin=*}
\usepackage{colortbl}
\usepackage{array} 
\usepackage{framed}
\DeclareUnicodeCharacter{2260}{$\neq$}

\makeatletter
\ifdefined\HCode
  \renewcommand{\rowcolors}[3]{}
\fi
\makeatother

\AtBeginDocument{%
  }

\ifpreprint
  \setcopyright{none}

  \renewcommand\footnotetextcopyrightpermission[1]{%
    \footnotetext{%
      \small\textit{%
      Preprint. This paper has been accepted to IUI ’26 (the 31st International Conference on Intelligent User Interfaces). The final published version will appear in the ACM Digital Library.}
    }%
  }
\else

\fi

\begin{document}

\title[Evaluating Generative AI in the Lab]{Evaluating Generative AI in the Lab: Methodological Challenges and Guidelines}

\author{Hyerim Park}
\orcid{0009-0006-4877-2255}
\authornote{The first two authors contributed equally to this research.}
\affiliation{%
  \institution{BMW Group}
  \city{Munich}
  \country{Germany}}
\affiliation{%
  \institution{University of Stuttgart}
  \city{Stuttgart}
  \country{Germany}}
\email{hyerim.park@bmw.de}

\author{Khanh Huynh}
\orcid{0009-0008-8330-147X}
\authornotemark[1]
\affiliation{%
  \institution{BMW Group}
  \city{Munich}
  \country{Germany}
}
\affiliation{%
  \institution{LMU Munich}
  \city{Munich}
  \country{Germany}}
\email{khanh.huynh@bmw.de}

\author{Malin Eiband}
\orcid{0000-0003-4024-1645}
\affiliation{%
  \institution{BMW Group}
  \city{Munich}
  \country{Germany}}
\email{malin.eiband@bmw.de}

\author{Jeremy Dillmann}
\orcid{0000-0002-5739-4704}
\affiliation{%
  \institution{BMW Group}
  \city{Munich}
  \country{Germany}}
\email{jeremy.dillmann@bmw.de}

\author{Sven Mayer}
\orcid{0000-0001-5462-8782}
\affiliation{%
  \institution{TU Dortmund University}
  \city{Dortmund}
  \country{Germany}}
\affiliation{%
  \institution{Research Center Trustworthy Data Science and Security}
  \city{Dortmund}
  \country{Germany}}
\email{info@sven-mayer.com}

\author{Michael Sedlmair}
\orcid{0000-0001-7048-9292}
\affiliation{%
  \institution{University of Stuttgart}
  \city{Stuttgart}
  \country{Germany}
}
\email{michael.sedlmair@visus.uni-stuttgart.de}

\renewcommand{\shortauthors}{Park et al.}

\begin{abstract}
Generative AI (GenAI) systems are inherently non-deterministic, producing varied outputs even for identical inputs. While this variability is central to their appeal, it challenges established HCI evaluation practices that typically assume consistent and predictable system behavior. Designing controlled lab studies under such conditions therefore remains a key methodological challenge. We present a reflective multi-case analysis of four lab-based user studies with GenAI-integrated prototypes, spanning conversational in-car assistant systems and image generation tools for design workflows. Through cross-case reflection and thematic analysis across all study phases, we identify five methodological challenges and propose eighteen practice-oriented recommendations, organized into five guidelines. These challenges represent methodological constructs that are either amplified, redefined, or newly introduced by GenAI's stochastic nature: (C1) reliance on familiar interaction patterns, (C2) fidelity--control trade-offs, (C3) feedback and trust, (C4) gaps in usability evaluation, and (C5) interpretive ambiguity between interface and system issues. Our guidelines address these challenges through strategies such as reframing onboarding to help participants manage unpredictability, extending evaluation with constructs such as trust and intent alignment, and logging system events, including hallucinations and latency, to support transparent analysis. This work contributes (1) a methodological reflection on how GenAI's stochastic nature unsettles lab-based HCI evaluation and (2) eighteen recommendations that help researchers design more transparent, robust, and comparable studies of GenAI systems in controlled settings.
\end{abstract}

\ccsdesc[500]{Human-centered computing~User studies}  
\ccsdesc[500]{Human-centered computing~HCI design and evaluation methods}  
\ccsdesc[500]{Human-centered computing~Empirical studies in HCI}  

\keywords{Generative AI, large language models (LLMs), user studies, methodology, human--AI interaction}

\maketitle

\section{Introduction}
Generative AI (GenAI) technologies are increasingly integrated into interactive systems across domains---from productivity tools to creative applications---by generating diverse forms of content such as text, images, and voice~\cite{shi_hci-centric_2024, almulla_understanding_2025, krakowski_human-ai_2025}. As these systems become more common in everyday contexts, evaluating their usability and user experience has become an important topic in HCI. However, GenAI also introduces challenges for established evaluation practices, particularly in controlled lab studies, which are among the core methods in HCI research~\cite{wolf_role_1989, kjeldskov_was_2014, rind_whys_2011}.

Unlike rule-based systems, such as traditional chatbots or in-car voice assistants, which produce fixed responses based on predefined commands, decision trees, or state machines~\cite{tan_rule-based_2025, silkhi_comparative_2024, singh_survey_2025}, GenAI models generate open-ended and context-dependent outputs that can differ with each interaction~\cite{zheng_evalignux_2025, ravera_usability_2025, simkute_ironies_2025, gmeiner_prototyping_2025, mackenzie_user_2015, preece_human-computer_1994}. This non-determinism, while enabling new forms of generative interaction, also disrupts key methodological assumptions that underlie lab-based evaluation, such as control, consistency, and comparability~\cite{pourasad_does_2025, kosch_risk_2024}. Without adapted approaches, researchers risk drawing misleading conclusions about usability, trust, or user behavior when stochastic system behavior is mistaken for interface design flaws. 
We do not claim that these challenges are unique to GenAI. Similar challenges can arise when evaluating other adaptive or intelligent systems~\cite{volkel_what_2020, todi_adapting_2021}. However, in generative systems, where output variability tends to be higher, these challenges are amplified and reframed, and additional challenges can emerge.
Designing studies that remain transparent, robust, and comparable under such stochastic conditions is thus an important step toward more reliable GenAI evaluation.

This unpredictability affects all stages of user studies, from task definition and prototype development to data collection and interpretation, yet how to plan and conduct such studies remains underexplored. Numerous GenAI studies focus on system performance or user-facing outcomes~\cite{shi_hci-centric_2024, yan_human-ai_2024, kolln_identifying_2022, park_we_2024, zamfirescu-pereira_why_2023}, while the methodological decisions and trade-offs behind evaluation design are rarely discussed. Some acknowledge issues such as hallucinations, shifting output styles, or novelty effects that influence trust and user experience~\cite{he_ai_2024, yun_user_2025}, yet these aspects are often mentioned only as study limitations rather than systematically examined through internal materials and reflections. Making such methodological reasoning explicit can help researchers anticipate challenges and design more rigorous evaluations of GenAI systems in controlled settings.

To address these issues, we conducted a \textbf{reflective multi-case study} of four controlled lab-based user studies involving GenAI-integrated prototypes. The cases span two domains---LLM-based conversational in-car assistants and GenAI image tools for professional design workflows---and vary in fidelity, modality, and participant groups. We adopted a multi-case approach because methodological challenges manifest differently across systems and study designs, and a single case would not capture this diversity or reveal recurring patterns~\cite{eisenhardt_building_1989}. Our goal was not to report user-facing outcomes such as task performance or satisfaction measures, but to analyze the methodological decisions, adaptations, and tensions that emerged throughout the research process. Drawing on study materials, researcher notes, and team discussions, we reflected on how GenAI's stochastic nature shaped study planning, prototyping, data collection, and analysis.
Our work addresses two research questions:
\begin{enumerate}
    \item What recurring methodological challenges arise when evaluating GenAI systems in controlled lab settings?
    \item How can these challenges be addressed in the design and execution of such studies?
\end{enumerate}

Through cross-case reflection, affinity diagramming, and inductive thematic analysis~\cite{braun_using_2006}, we identified \textbf{five methodological challenges (C1--C5)} that complicate established HCI evaluation practices. These challenges represent methodological constructs that are either \textbf{amplified, redefined, or newly introduced} by the generative and non-deterministic nature of GenAI systems: (C1) amplified reliance on familiar interaction patterns, (C2) amplified trade-offs between fidelity and experimental control, (C3) redefined feedback loops and user trust, (C4) new methodological gaps in usability evaluation, and (C5) amplified interpretive ambiguity between interface and system behavior.
Building on these findings, we propose \textbf{five methodological guidelines (G1--G5)}, each linked to one of the challenges, and eighteen practice-oriented recommendations that offer actionable strategies for designing, conducting, and analyzing GenAI user studies. The guidelines include preparing participants for unpredictable system behavior, aligning prototype fidelity to study goals, improving feedback interpretability and user trust, adapting evaluation strategies to capture GenAI-specific experiences, and building flexibility into study design and analysis.

This paper contributes (1) a methodological reflection based on four GenAI-integrated lab studies that reveal how stochastic model behavior challenges established evaluation practices, and (2) eighteen concrete recommendations, structured under five guidelines, to support the planning and execution of GenAI user studies in controlled research settings.

\section{Related Work}
User studies are a foundational method in HCI for evaluating interactive systems~\cite{mackenzie_user_2015, lazar_research_2010}. By typically combining quantitative measures (e.g., surveys, task logs) with qualitative feedback data, user studies help assess usability, user experience, and task performance across diverse interface types~\cite{preece_human-computer_1994, schmidt_evaluation_2021, greenberg_usability_2008}. Traditional evaluations often rely on controlled lab experiments, measuring completion time, error rates, or subjective satisfaction, and are well-suited to systems with well-defined tasks and stable behavior~\cite{preece_human-computer_1994, greenberg_usability_2008, lazar_research_2010}.

\subsection{Methodological Shifts in HCI Evaluation}
As interactive systems become more adaptive, open-ended, and embedded in dynamic contexts, conventional evaluation methods such as usability testing and short-term lab studies often prove insufficient. Poppe et al.~\cite{poppe_evaluating_2007} emphasize that systems involving novel sensing technologies and shifting user/system initiative require longitudinal observations and context-sensitive evaluation.
Similarly, Greenberg et al.~\cite{greenberg_usability_2008} argue that standardized usability testing may hinder innovation or overlook the critical experiential dimensions, especially in systems supporting exploration, creativity, or collaboration. 
Brdnik et al.~\cite{brdnik_intelligent_2022}, in a review of IUI papers from 2012 to 2022, found that many evaluations still rely on conventional experiments and questionnaires, with limited attention to metrics suited for system adaptivity or the dynamics of human-AI co-adaption. Similar limitations have been observed in newer interaction paradigms such as voice interfaces, AR/VR, and IoT~\cite{capdevila_current_2021}.

In response, the HCI community has adopted a range of complementary methods, including in-the-wild studies~\cite{rogers_research_2017}, longitudinal deployments~\cite{kjaerup_longitudinal_2021}, Wizard-of-Oz studies~\cite{klemmer_suede_2000}, simulation and modeling approaches~\cite{murray-smith_what_2022}, and more interpretive, mixed-method designs that combine usage data with reflective user feedback~\cite{creswell_designing_2017}. These approaches are often applied to systems that are complex, adaptive, or open-ended, or that are embedded in real-world settings where traditional lab evaluations may be insufficient. 

The information visualization community provides a documented and explicit example of methodological evolution in response to similar challenges. Visualization tools are often used for exploratory tasks like analyzing large datasets or generating insights where no single correct answer or predefined success criterion exists~\cite{plaisant_challenge_2004, lam_empirical_2012}. In such contexts, traditional metrics like task completion time and error rates may misrepresent how users interact or derive value from the system over time. To address these challenges, the BELIV (Beyond Time and Errors) workshop series~\cite{bertini_beliv08_2007}\footnote{\href{https://beliv-workshop.github.io/}{https://beliv-workshop.github.io/}} was established to promote evaluation approaches designed for the unique challenges faced by visualization systems, emphasizing user engagement, exploration, and sense making beyond traditional performance metrics.
Building on this foundation, researchers have proposed structured evaluation frameworks. \citet{lam_empirical_2012} introduced a taxonomy of seven evaluation scenarios to guide method selection based on system type and research goal. Munzner's nested model~\cite{munzner_nested_2009} defines multiple design levels, such as data abstraction, visual encoding, interaction techniques, and domain tasks, clarifying what is being evaluated and how. \citet{sedlmair_design_2012} reflect on the practical challenges of real-world design studies, emphasizing the ``messiness'' and need for context-specific adaptation when evaluating visualization systems outside controlled lab settings.  
Together, these methodological considerations highlight challenges that arise when evaluating complex, adaptive, or user-driven systems in HCI~\cite{sedlmair_design_2012, dalsgaard_reflective_2012}. GenAI shares similar characteristics: its outputs are inherently variable and context-dependent, making them difficult to evaluate using fixed metrics or predefined task goals.

Such methodological shifts have also emerged during periods of disruption or modality-specific innovation. For example, \citet{schmidt_evaluation_2021} explored remote and out-of-the-lab evaluation strategies in response to pandemic-related constraints, proposing alternatives to in-person testing such as browser-based prototypes and the reuse of existing datasets. Similarly, new interaction modalities have prompted adaptations in evaluation practice. Voice user interfaces (VUIs), for instance, prompted new methods focused on conversational timing, speech clarity, and context-aware interaction~\cite{kolln_identifying_2022}. Tools such as HEUROBOX were developed to identify voice-specific usability issues~\cite{fulfagar_development_2021}, and standard instruments like the System Usability Scale (SUS) were adapted to assess qualities such as naturalness, politeness, and conversational flow~\cite{iniguez-carrillo_usability_2021}. These efforts reflect a broader push toward inclusive and context-sensitive evaluation strategies for emerging interaction paradigms~\cite{stigall_systematic_2020}.

Despite this progress, limited attention has been paid to the methodological implications of evaluating non-deterministic GenAI systems in controlled lab settings. Our work addresses this gap by reflecting on researcher decisions, trade-offs, and adaptations in the context of user studies involving GenAI.

\subsection{Evaluating GenAI Systems in HCI}
GenAI systems, including LLMs and image generators, introduce distinct methodological challenges for HCI evaluation. Unlike deterministic systems, GenAI tools are inherently variable, producing different outputs for the same input prompt, even under similar conditions~\cite{brown_language_2020, simonen_exploration_2025, shi_hci-centric_2024}. This variability complicates evaluation in multiple ways. Traditional metrics, such as task completion time, accuracy, and error rates, rely on consistent system behavior and clear success criteria; yet, many GenAI use cases---including creative generation, idea exploration, and open-ended problem solving---lack a single correct outcome. Moreover, responses may vary in quality, structure, length, and completeness, making comparisons across participants difficult. GenAI systems are also known to produce hallucinations, that is, outputs that appear plausible but are factually incorrect. This can undermine output-based measures of reliability or satisfaction, particularly when users initially trust the response.

Prior work in HCI and explainable AI (XAI) has explored user study methodologies for evaluating AI systems. For instance, Rong et al.~\cite{rong_towards_2024} review how XAI research primarily focuses on interpreting decision boundaries and building user trust in deterministic systems, often centered on decision-support tools or classifiers with predictable behaviors. In contrast, GenAI systems produce open-ended, multi-modal, and variable outputs, introducing unpredictability and interpretive ambiguity that are not central in XAI. Interactions with GenAI systems are typically iterative and involve a co-construction of meaning between user and system, with user expectations and satisfaction shaped by subjective, context-dependent factors. Together, these differences present challenges that call for rethinking user study design in HCI.
 
Existing user studies investigating GenAI systems also highlight several distinct challenges. While GenAI systems excel at generating general content, they often struggle with domain-specific understanding and fine-grained detail. As a result, outputs tend to be more generalized, lacking the depth and expertise required in specific fields~\cite{yan_human-ai_2024, he_ai_2024, lee_storagechat_2025, su_enhancing_2024}.
Prior work further shows that participants' mental models of and skepticism towards GenAI frequently shape their responses in user studies. Participants often draw on their prior experiences with GenAI---both positive and negative---when forming opinions during studies. A recurring concern is GenAI's tendency to produce seemingly plausible but factually incorrect content~\cite{yan_human-ai_2024, he_ai_2024}. 
In addition, reliance on AI can introduce cognitive biases, such as confirmation bias or the uncritical acceptance of agreeable responses~\cite{yun_generative_2025, he_ai_2024, yun_user_2025}. Interviews have also reported novelty effects associated with so-called ``advanced'' AI tools, introducing another source of bias. Despite recognizing these issues, many studies acknowledge them only as limitations and fall short of proposing comprehensive methodological responses. 
To address these challenges, researchers have employed alternative approaches in GenAI studies. One common strategy is to use pre-generated outputs to control variability and ensure consistency. Other studies adopt Wizard of Oz methodologies, in which human facilitators simulate GenAI capabilities in real-time. However, this approach introduces its own limitations, such as delays in response and gaps in domain expertise~\cite{rayan_exploring_2024}. Overall, most prior work emphasizes real-world usage scenarios and treats unpredictability and non-determinism primarily as limitations, rather than examining their implications for lab-based evaluation of generative systems.

Beyond GenAI-specific work, our contribution builds on a broader body of HCI research examining how evaluation methods need to be adapted for systems with uncertain, opaque, or partially simulated behavior. Wizard-of-Oz studies have long highlighted trade-offs between experimental control, realism, and interpretability, as well as challenges of attribution when system behavior is mediated by human or hybrid components~\cite{riek_wizard_2012, steinfeld_oz_2009, dahlback_wizard_1993}. More recent work has extended these concerns to machine learning (ML)--enabled systems, demonstrating that realistically simulating ML errors in Wizard-of-Oz studies is itself methodologically challenging and has consequential effects on user experience evaluation~\cite{jansen_wizard_2022}. Additionally, critiques of usability evaluation methods argue against viewing them as ``fixed recipes,'' urging the adaptation of methodological resources to align with specific system properties and research goals~\cite{woolrych_ingredients_2011}. Building on these perspectives, we present guidelines as modular methodological resources for designing and interpreting lab-based studies of GenAI systems.

\section{Methodology: A Multi-Case Study with Reflection and Thematic Analysis}
We adopted a \textbf{multi-case study approach} supported by reflective and inductive thematic analysis to examine methodological challenges in evaluating GenAI systems through lab-based user studies. This approach enabled us to capture methodological challenges across varied GenAI systems and study designs while reflecting on our research decisions. A multi-case perspective supported the identification of both recurring patterns and context-specific nuances across studies~\cite{gustafsson_single_2017}. By combining Eisenhardt's \textbf{systematic multi-case framework}~\cite{eisenhardt_building_1989} with \textbf{affinity diagramming}~\cite{beyer_contextual_1999} for initial structuring and \textbf{inductive thematic analysis}~\cite{braun_reflecting_2019, cooper_thematic_2012}, we moved from concrete study observations to broader methodological insights. The resulting five challenges (C1--C5) represent \textbf{recurring methodological tensions} in lab-based GenAI evaluation and form the foundation for the guidelines presented in \autoref{sec:guidelines}. In summary, this analytic process combined affinity diagramming for initial structuring with thematic analysis to synthesize cross-case methodological challenges.

\subsection{Case Selection and Context}
We analyzed four user studies conducted between \textit{January 2024} and \textit{September 2025}, covering both conversational and visual GenAI systems. We selected these cases to support cross-case comparison, following established guidance for multi-case research~\cite{eisenhardt_building_1989}, rather than to maximize the number of challenges identified. Accordingly, our inclusion criteria were guided by methodological considerations and practical access constraints. We included only lab-based user studies that we conducted ourselves, involved direct user interaction with a GenAI system, and empirically evaluated that interaction with participants. Because all authors were involved in the studies, we had detailed insight into design rationale, trade-offs, and study-planning decisions, supporting retrospective methodological analysis.
The cases varied in system type, participant group, and prototype fidelity---from in-car voice assistants to design-oriented image generation tools---providing sufficient diversity to examine recurring methodological challenges across contexts. This scale aligns with established guidance for multi-case research, which suggests that a small number of heterogeneous cases can provide depth and analytical comparability~\cite{eisenhardt_building_1989, gustafsson_single_2017}. 
As such, the resulting challenges should be understood as illustrative rather than exhaustive, highlighting recurring methodological tensions rather than providing a complete landscape of GenAI evaluation issues. 
We approached reflection as a method for methodological inquiry rather than subjective introspection, triangulating researcher memos, study logs, and artifacts to ensure transparency. Each case featured different combinations of GenAI models, interaction modalities, user goals, and study designs. This diversity offered a comparative basis for identifying recurring methodological tensions across distinct study contexts. An overview of the cases is provided in \autoref{tab:case_study_overview}, with detailed descriptions in \autoref{app:cases}.

\begin{table*}[t]
    \centering
    \setlength{\tabcolsep}{6pt} 
    \renewcommand{\arraystretch}{1.1} 
    \caption{Summary of the four lab-based studies highlighting study context, user groups, primary interaction modalities, adopted GenAI models, prototype fidelity, evaluation methods, and study timeframe.}
    \label{tab:case_study_overview}
    \Description{The table compares four lab-based case studies of GenAI systems labeled Case A through Case D. Columns correspond to individual studies, and rows summarize key study aspects including study overview, user groups, interaction modalities, AI models used, prototype type and fidelity, evaluation methods, and study timeframe. The table provides a side-by-side overview of how the studies differ in context, participants, system implementations, and evaluation approaches.}
    \footnotesize

    \rowcolors{2}{gray!7}{white}
    \begin{tabularx}{\linewidth}{>{\raggedright\arraybackslash}p{2.3cm}XXXX}
    \toprule
    \textbf{Study Aspect} & \textbf{Case A} & \textbf{Case B} & \textbf{Case C} & \textbf{Case D} \\ 
    \midrule
    
    \textbf{Study Overview} &
    A multimodal conversational in-car assistant powered by an LLM, exploring interactions across different driving-related use cases. &
    A multimodal conversational LLM-based in-car assistant with integrated GUI interaction, focusing on users' references to visual elements. &
    A paper-based prototype of an early-stage GenAI image generation tool, targeting professional designers' input strategies. &
    A fully functional GenAI image generation tool deployed in professional design practice to support interactive workflows. \\
    \addlinespace
    
    \textbf{User Groups} &
    Drivers &
    Drivers and passengers &
    Professional designers &
    Professional designers and design students \\
    \addlinespace
     
    \textbf{Interaction Modalities} &
    Voice and GUI interaction (touchscreen) &
    Voice and GUI interaction (touchscreen) &
    Text input, scribbling, and handwritten annotations &
    Text input and visual inputs (scribbling and annotations), or combinations of both \\
    \addlinespace   
    
    \textbf{AI Models Used} &
    GPT-4o &
    GPT-4o &
    DALL·E 2 &
    DALL·E 3, GPT-image-1, Flux.1 Kontext Pro, GPT-4o \\
    \addlinespace
    
    \textbf{Prototype Type and Fidelity} &
    Fully functional &
    Fully functional &
    Paper-based prototype &
    Fully functional \\
    \addlinespace
     
    \textbf{Evaluation Methods} &
    Semi-structured interviews and usability surveys (Likert scale) &
    Semi-structured interviews and usability surveys (Likert scale) &
    Interviews and think-aloud methods &
    Comparative study, interviews, and custom surveys (Likert scale) \\
    \addlinespace
     
    \textbf{Study Timeframe} &
    January 2024--April 2024 &
    March 2024--August 2024 &
    August 2024--January 2025 &
    March 2025--September 2025 \\
    
    \bottomrule
    \end{tabularx}
\end{table*}

\subsection{Data Collection}
We collected a range of internal research materials, including study plans, interview guides, prototype specifications, and observational notes from user sessions. Researcher reflections and memos written during and after each study, together with meeting summaries and study logs, were used to capture the reasoning behind methodological decisions. These materials enabled us to reconstruct both the practical procedures and the rationale behind specific design choices. Each researcher independently identified methodological issues from the studies they primarily planned and conducted. All observations and notes were then consolidated into a shared Figma workspace, where we collaboratively externalized, organized, and discussed emerging methodological patterns. To support collaborative organization of the collected materials, we employed \textbf{affinity mapping techniques}~\cite{beyer_contextual_1999} to iteratively group, split, and reorganize methodological observations through team discussion.

\subsection{Analysis}
Our analysis followed an inductive thematic analysis applied across the four lab-based studies. Rather than starting from predefined categories, we iteratively developed themes grounded in the collected data. The analysis builds on materials generated through an initial affinity diagramming step, which supported the organization and externalization of methodological observations before formal thematic analysis. This process unfolded in three stages: (1) organizing methodological observations within each case, (2) collaboratively clustering and comparing patterns across cases, and (3) synthesizing broader methodological challenges emerging from the analysis~\cite{eisenhardt_building_1989}.

\paragraph{Step 1: Organizing Case Observations.}
In the first stage, we organized all methodological observations and reflections according to study phases (e.g., research planning, prototyping, participant interaction, data collection, and analysis), supported by affinity diagramming. This phase-based coding allowed us to identify when and where methodological challenges occurred within the study process and supported early visualization of emerging patterns across projects. \autoref{app:initial_analysis} shows this early phase-based clustering.

\paragraph{Step 2: Collaborative Coding and Clustering.}
Next, building on this initial organization, we conducted \textbf{collaborative coding and clustering} to identify recurring methodological patterns. Using the materials generated through affinity diagramming, the authors iteratively compared and merged related analytic codes while discussing conceptual relationships across studies. This process moved from detailed, case-specific observations (codes) to conceptual clusters (subthemes) that captured methodological issues recurring across multiple contexts.

\begin{figure*}[t]
    \centering
    \includegraphics[width=\linewidth]{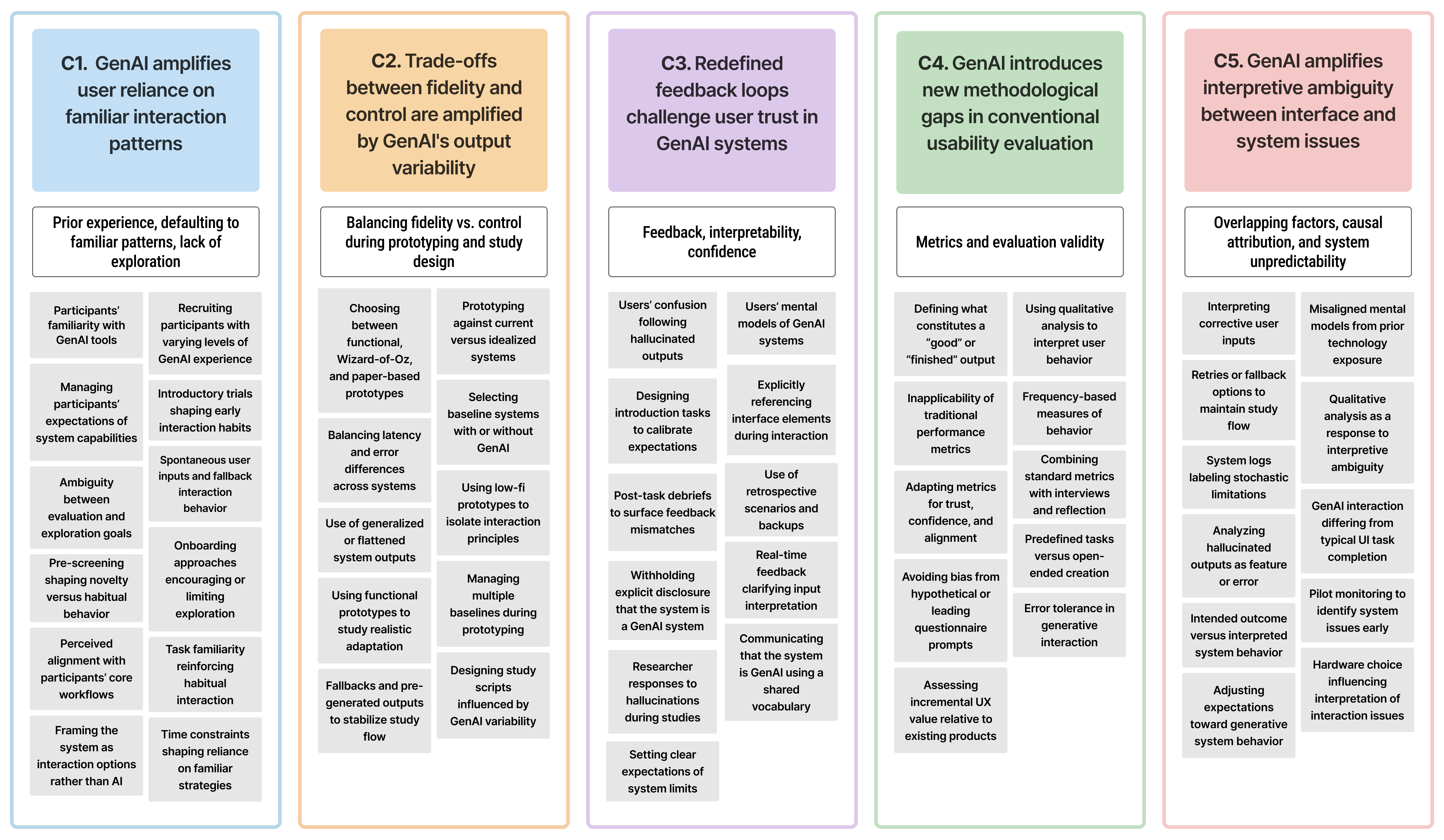}
    \caption{Visualization of our inductive thematic coding process from initial analytic codes (light gray boxes) to subthemes (white boxes) and five higher-order themes (C1--C5). Each column represents one methodological challenge that emerged through iterative comparison and clustering across studies. Detailed documentation of affinity notes, analytic codes, subthemes, themes, and methodological reflections is provided in the supplementary material.}
    \label{fig:coding}
    \Description{The figure shows a five-column diagram representing an inductive thematic coding process for studies of GenAI systems. Each column corresponds to one higher-order methodological challenge (C1--C5) and is labeled at the top. Within each column, white boxes denote subthemes, and light gray boxes list the initial analytic codes that informed them. The diagram illustrates how analytic codes were clustered into subthemes and synthesized into five challenges related to reliance on familiar interaction patterns, amplified fidelity--control trade-offs, altered feedback loops affecting user trust, gaps in usability evaluation methods, and ambiguity in attributing behavior to interface--system boundaries.}
\end{figure*}

\paragraph{Step 3: Thematic Synthesis.}
Finally, through \textbf{thematic synthesis}, we abstracted these conceptual clusters into five \textbf{higher-level methodological challenges (C1--C5)}. This synthesis combined descriptive coding with reflective interpretation, focusing on methodological tensions we encountered repeatedly, such as participant familiarity, prototype fidelity, interpretability of feedback, metric validity, and confounding system factors. \autoref{fig:coding} shows this progression from initial analytic codes (specific observations) to focused subthemes (conceptually related methodological issues) and five higher-order themes (challenges).

\section{Results: Challenges Identified Across Case Studies} 
\label{sec:challenges}
Our inductive thematic analysis of the four lab-based GenAI studies revealed \textbf{five methodological challenges (C1--C5)} that complicate conventional HCI evaluation practices. These challenges reflect methodological constructs that are either \textbf{amplified, redefined, or newly introduced} by the generative and non-deterministic nature of GenAI systems. They emerged through iterative comparison of authors' reflections, study artifacts, and participant observations, capturing tensions spanning study planning, prototyping, user interaction, evaluation, and interpretation. Together, they illustrate how GenAI systems reshape established assumptions about what can be controlled, measured, and meaningfully interpreted in lab research. \autoref{fig:challenges_x_recommendations} provides an overview of the core challenges and the corresponding methodological recommendations. In the following, we use \emph{output variability} to describe the degree to which a system may produce different outputs for the same input under similar conditions---it is stochastic, but not arbitrary or completely unpredictable.

\subsection{C1. GenAI amplifies user reliance on familiar interaction patterns}
\label{sec:c1}
Users in HCI studies often default to familiar interaction strategies when introduced to novel systems. In GenAI contexts, this tendency becomes more pronounced, not merely due to habit or bias, but because the system's stochastic feedback prevents stable learning. Without predictable input--output mappings---as found in conventional interfaces that follow rule-based or deterministic logic---users struggle to infer how the system interprets their actions and instead rely on previously learned strategies, such as favoring text over visual input in GenAI image tools or conventional phrasing in voice commands. Across our GenAI cases, this reliance was reinforced by inconsistent responsiveness: participants preferred input modes that seemed more legible to the system, even when novel modalities were available.
For example, participants in \textbf{Cases A and B} often relied on their expectations of how in-car voice assistants typically behave, referring to GUI elements as if the systems followed deterministic interaction rules. Similarly, in \textbf{Cases C and D}, some designers preferred text prompts over sketches or annotations, citing prior training and confidence in text-based interaction. This reliance limited participants' exploration of new affordances and, consequently, the study's ability to evaluate novel interaction designs.

\textbf{Takeaway:} GenAI's unpredictability \textbf{amplifies} users' reliance on familiar input modes, reducing their willingness to explore new affordances and constraining what lab studies can reveal about interaction designs.

\subsection{C2. Trade-offs between fidelity and control are amplified by GenAI's output variability}
\label{sec:c2}
Interactive system studies often aim to balance \textbf{experimental control} with \textbf{ecological realism}---a long-standing methodological tension in HCI. In GenAI evaluations, however, this balance becomes especially difficult because stochastic outputs introduce additional, uncontrollable variability on top of existing system behavior. Low-fidelity setups (e.g., scripted responses or Wizard-of-Oz methods) can increase consistency but may limit the generative qualities that characterize GenAI, while high-fidelity prototypes capture more authentic behavior but can introduce noise and unpredictability that complicate study outcomes.
In \textbf{Case D}, low-fidelity prototypes increased consistency but limited opportunities for exploration. Conversely, in \textbf{Cases A, B, and D}, fully functional prototypes enabled genuine generative interactions yet led to latency and unexpected responses. Participants expressed confusion when outputs deviated from expectations, despite prior briefings about system limitations.

\textbf{Takeaway:} This challenge extends a familiar HCI problem---balancing fidelity and control---but GenAI's higher degree of output variability \textbf{amplifies} this impact, making design choices methodological in nature, as they directly affect study reliability and determine which types of interactions can be meaningfully observed.

\subsection{C3. Redefined feedback loops challenge user trust in GenAI systems}
\label{sec:c3}
In conventional interfaces, feedback mechanisms are typically designed to be interpretable, enabling users to infer the relationship between their input and the resulting output. GenAI systems complicate this loop, as stochastic generation and opaque model behavior make it unclear how inputs are processed or acknowledged. Across our studies, participants frequently expressed doubt about whether the system recognized or understood their input, especially when using less familiar modalities (e.g., voice in the car study, scribbles in the design study).
In \textbf{Case B}, conversational hallucinations led participants to question whether miscommunication stemmed from their phrasing or from the model's unpredictable generation of irrelevant responses. In \textbf{Case D}, scribble-based prompts sometimes yielded mismatched or irrelevant images, leading participants to wonder whether the system had misunderstood their visual input or failed to align it with the intended concept. In contrast, text prompts provided more consistent and traceable responses, which participants perceived as more reliable.

\textbf{Takeaway:} In GenAI systems, feedback is not simply less reliable but \textbf{redefined}. Users may hesitate to explore unfamiliar modalities not because of interface flaws, but because feedback can be unstable or ambiguous, which can disrupt trust and engagement during the study itself. This instability can limit the extent to which evaluators can reliably infer from observed interaction behavior.

\subsection{C4. GenAI introduces new methodological gaps in conventional usability evaluation} 
\label{sec:c4}
Standard usability measures such as the System Usability Scale (SUS) or the User Experience Questionnaire (UEQ) were developed under the assumption of systems that produce relatively consistent and reproducible responses. These instruments assume clear input–output relationships and stable performance, allowing numerical scores to reflect interface usability. GenAI systems violate these assumptions: stochastic and context-dependent outputs mean that user ratings often capture variability in system behavior rather than interface design quality.
In \textbf{Case B}, SUS ratings were frequently influenced by factors such as hallucinations, latency, or unexpected output variation rather than interface design, as indicated by follow-up interviews. Similarly, in \textbf{Case D}, low UEQ scores for scribble-based tasks were sometimes attributed to confusing or incoherent outputs rather than to interaction design flaws. Such findings suggest that established usability metrics may not reliably separate interface-related issues from variability arising from generative system behavior. To support a valid interpretation, qualitative observations and mixed methods (e.g., interviews, think-aloud protocols) were essential for contextualizing numerical results.

\textbf{Takeaway:} GenAI creates a \textbf{new methodological gap} in usability evaluation. Traditional metrics assume consistency and determinism, yet user frustration often arises from model unpredictability rather than interface design, highlighting the need for new or hybrid evaluation approaches that account for stochastic behavior.

\subsection{C5. GenAI amplifies interpretive ambiguity between interface and system issues}
\label{sec:c5}
In conventional systems, usability breakdowns can often be traced to specific causes, such as interface design issues or users’ interaction mistakes. In GenAI studies, however, overlapping factors---interface affordances, user strategies, and model behavior---make it challenging to determine the origin of observed problems. Unlike the interaction-level uncertainty described in \autoref{sec:c3}, this challenge instead arises during \textbf{post-hoc interpretation}, when researchers attempt to attribute causes to observed outcomes. Across our cases, participant confusion or task failures often stemmed from intertwined system- and interface-level factors, blurring evaluative judgment.
This ambiguity is further amplified when variability in system output makes it harder to attribute observed behavior to interface design decisions rather than underlying model behavior.
In \textbf{Case D}, irrelevant image outputs from scribbles could signal either poor interface affordances or stochastic model behavior. Similarly, in \textbf{Case B}, hallucinated or delayed responses disrupted task flow, making it unclear whether confusion arose from design limitations or the model's unpredictable processing. Without precise system logging or detailed observational notes marking critical events (which tend to occur more frequently in GenAI systems than in conventional ones), tracing the source of such breakdowns and interpreting their methodological implications reliably becomes more difficult.

\textbf{Takeaway:} GenAI's unpredictability \textbf{amplifies} existing interpretive challenges in usability evaluation. Overlapping effects from user behavior, interface design, and model responses blur causal boundaries, making it harder to determine whether observed issues reflect design flaws or generative variability.

\section{Guidelines for Designing and Evaluating GenAI Systems in Controlled Studies}
\label{sec:guidelines}
\begin{figure*}[t]
    \centering
    \includegraphics[width=\linewidth]{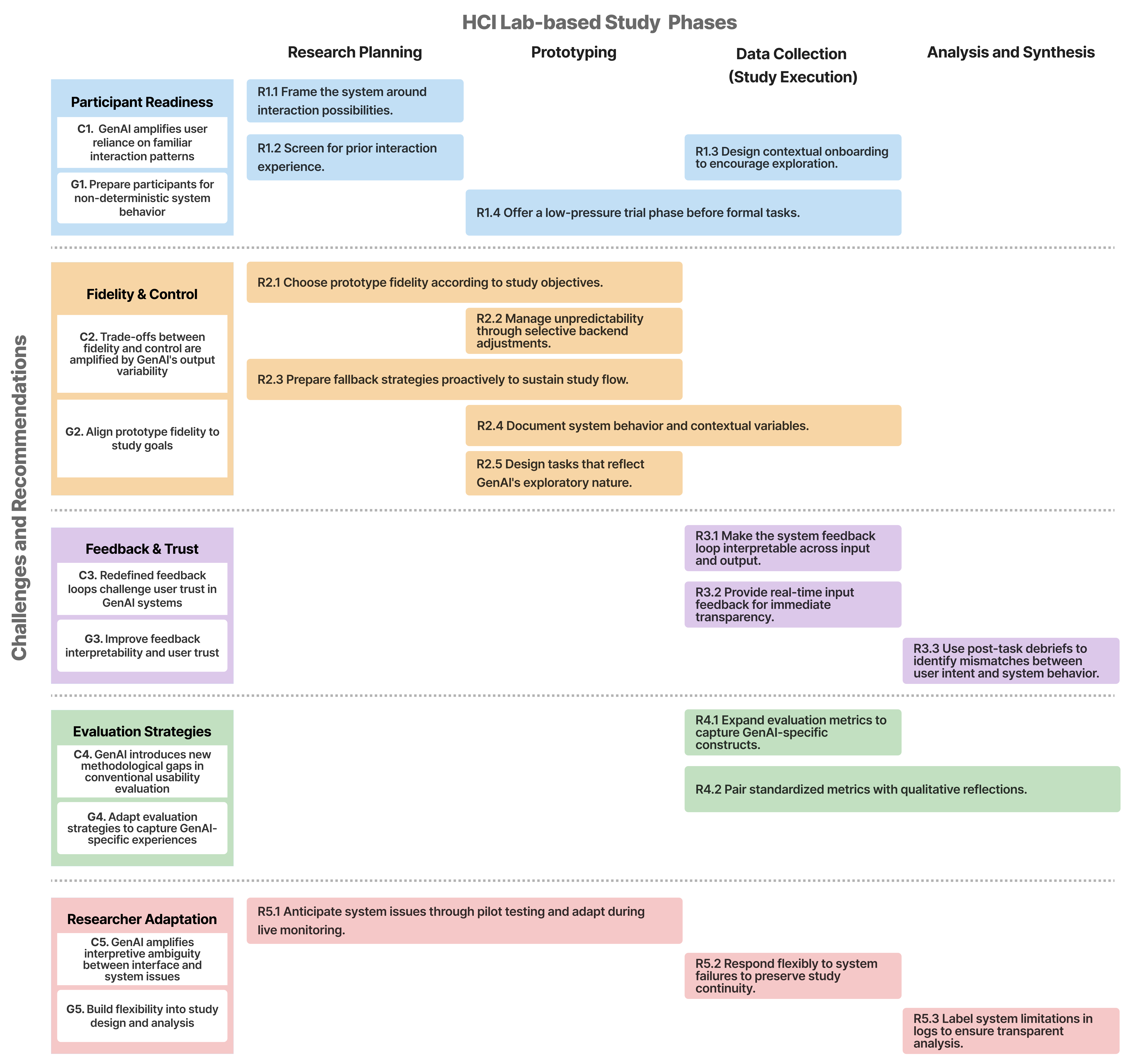}
    \caption{Methodological challenges (C1--C5) and corresponding guidelines (G1--G5) with all eighteen recommendations (R1.1--R5.3) for evaluating GenAI systems in HCI lab studies. The figure illustrates where each recommendation primarily applies across research phases, from planning and prototyping to data collection and analysis.}
    \label{fig:challenges_x_recommendations}
    \Description{The figure is a matrix-style diagram mapping methodological challenges and recommendations for evaluating GenAI systems in HCI lab studies across research phases. The horizontal axis shows four phases---research planning, prototyping, data collection, and analysis and synthesis. The vertical axis groups five challenges (C1--C5), each paired with a corresponding guideline (G1--G5): participant readiness, fidelity and control, feedback and trust, evaluation strategies, and researcher adaptation. Within each row, colored boxes labeled R1.1 through R5.3 indicate individual recommendations and their primary phase of application. The figure shows how eighteen recommendations are distributed across study phases, with some challenges spanning multiple phases and others concentrated in execution or analysis.}
\end{figure*}

Building on the five methodological challenges (C1--C5) identified in controlled lab studies, we developed \textbf{five methodological guidelines (G1--G5)}, each accompanied by a set of \textbf{practice-oriented recommendations}. While the challenges highlight how GenAI complicates established evaluation practices, the guidelines provide actionable strategies for designing, conducting, and analyzing GenAI lab studies more effectively. Each guideline responds to a core methodological issue, including preparing participants for stochastic systems (G1), balancing control and fidelity (G2), improving feedback interpretability (G3), adapting usability metrics (G4), and strengthening post-hoc interpretation (G5). They aim to help HCI and UX researchers anticipate, document, and mitigate the unique methodological tensions introduced by GenAI's non-deterministic behavior. \autoref{tab:recommendations_summary} summarizes the five guidelines and their corresponding recommendations, while \autoref{fig:challenges_x_recommendations} illustrates how these guidelines relate to the five challenges across study phases.

\begin{table*}[t]
    \centering
    \setlength{\tabcolsep}{8pt} 
    \renewcommand{\arraystretch}{1.4} 
    \caption{Overview of the five methodological guidelines and their associated recommendations, showing how they address core challenges in evaluating GenAI systems.}
    \label{tab:recommendations_summary}
    \Description{The table summarizes five methodological guidelines for evaluating GenAI systems and their associated recommendations. Rows correspond to guidelines G1 through G5, each labeled with a short descriptor indicating the related challenge area. The first column lists the guideline identifier, the second column provides the guideline description, and the third column enumerates the specific recommendations (R1.1--R5.3) associated with each guideline. The table shows how multiple recommendations are grouped under each guideline to address different aspects of participant readiness, prototype fidelity, feedback and trust, evaluation strategies, and researcher adaptation.}
    \small
    
    \rowcolors{2}{gray!7}{white}
    \begin{tabularx}{\linewidth}{p{0.4cm} >{\raggedright\arraybackslash}p{5.4cm} >{\raggedright\arraybackslash}X}
    \toprule
    \textbf{G\#} & \textbf{Guideline} & \textbf{Key Recommendations} \\
    \midrule
    
    G1 & Prepare participants for non-deterministic system behavior {\footnotesize\textit{(Participant readiness)}} 
    & {\setlength{\parskip}{3pt}%
    \textbf{R1.1} Frame the system around interaction possibilities \par
    \textbf{R1.2} Screen for prior interaction experience \par
    \textbf{R1.3} Design contextual onboarding to promote exploration \par
    \textbf{R1.4} Offer a low-pressure trial phase before formal tasks
    }\\
    \addlinespace[2pt]
    
    G2 & Align prototype fidelity to study goals {\footnotesize\textit{(Fidelity \& control)}} 
    & {\setlength{\parskip}{3pt}%
    \textbf{R2.1} Choose prototype fidelity according to study objectives \par
    \textbf{R2.2} Manage unpredictability through selective backend adjustments \par
    \textbf{R2.3} Prepare fallback strategies proactively to sustain study flow \par
    \textbf{R2.4} Document system behavior and contextual variables \par
    \textbf{R2.5} Design tasks that reflect GenAI's exploratory nature 
    } \\
    \addlinespace[2pt]
    
    G3 & Improve feedback interpretability and user trust {\footnotesize\textit{(Feedback \& trust)}} 
    & {\setlength{\parskip}{3pt}%
    \textbf{R3.1} Make the system feedback loop interpretable across input and output \par
    \textbf{R3.2} Provide real-time input feedback for immediate transparency \par
    \textbf{R3.3} Use post-task debriefs to identify mismatches between user intent and system behavior 
      } \\
    \addlinespace[2pt]
    
    G4 & Adapt evaluation strategies to capture GenAI-specific experiences {\footnotesize\textit{(Evaluation strategies)}} 
    & {\setlength{\parskip}{3pt}%
    \textbf{R4.1} Expand evaluation metrics to capture GenAI-specific constructs \par
    \textbf{R4.2} Pair standardized metrics with qualitative reflections
      } \\
    \addlinespace[2pt]
    
    G5 & Build flexibility into GenAI study design and analysis {\footnotesize\textit{(Researcher adaptation)}} 
    & {\setlength{\parskip}{3pt}%
    \textbf{R5.1} Anticipate system issues through pilot testing and live monitoring \par
    \textbf{R5.2} Respond flexibly to system failures to preserve study continuity \par
    \textbf{R5.3} Label system limitations in logs to ensure transparent analysis    
    } \\
    \bottomrule
    \end{tabularx}
\end{table*}

\subsection{G1. Prepare Participants for Non-Deterministic System Behavior}  
Participants often approach GenAI systems with expectations shaped by prior experiences with deterministic interfaces (e.g., voice assistants or design tools). When outputs vary across identical inputs, these expectations can lead to confusion, frustration, or misattribution of errors to the interface.
Preparing participants for stochastic behavior helps them interpret variability as an inherent property of the system rather than as a failure. The four recommendations under G1 address participant preparation as a gradual process, starting with shaping expectations during system introduction (R1.1), accounting for prior experience (R1.2), guiding initial interactions through structured onboarding (R1.3), and concluding with free exploration in a low-pressure setting (R1.4).

\paragraph{R1.1 Frame the system around interaction possibilities.}
How a system is introduced influences how participants approach it.
Framing it as ``AI-powered'' or referencing commercial tools (e.g., ChatGPT, Midjourney) can narrow participants' expectations and lead them to rely on familiar input patterns even when other modalities are available.
To encourage broader exploration, present the system in terms of its \textbf{input possibilities, supported tasks, and limitations} rather than emphasizing its AI identity.
In \textbf{Case C and D}, where the system was introduced as an AI tool, users often defaulted to text prompts despite having access to scribbles and annotations, reflecting prior experience with text-based GenAI tools that participants cited when reasoning about their interaction strategies. In contrast, in \textbf{Cases A and B}, the systems were presented through their task functions, and information about the underlying LLMs was disclosed only after the study, which appeared to promote more varied and open-ended interaction.

\paragraph{R1.2 Screen for prior interaction experience.}
Participants' previous exposure to specific input modalities can shape both their confidence and their evaluation of usability. Without accounting for this, it becomes difficult to distinguish genuine usability issues from those arising from \textbf{unfamiliarity} (e.g., hesitation or confusion) or \textbf{over-familiarity} (e.g., reluctance to explore alternative inputs).
We therefore recommend \textbf{pre-screening participants} for prior experience with modalities such as in-car voice assistants, stylus input, or prompt-based systems to help support more accurate interpretation of observed behaviors. In \textbf{Case B}, the absence of such screening meant that participants defaulted to typical home-assistant commands when they were uncertain how to proceed. In contrast, \textbf{Case A} recruited experienced in-car voice assistant users, which enabled clearer differentiation between skill-based challenges and system-related issues.
Familiarity also shaped exploration. In \textbf{Cases C and D}, participants with strong prior experience using text-based GenAI tools often defaulted to text prompts, while those without prior stylus experience struggled initially. Together, these familiarity effects complicated attribution, making it difficult to determine whether performance issues stemmed from the input method itself, habitual interaction patterns, or device unfamiliarity.

\paragraph{R1.3 Design contextual onboarding to encourage exploration.}
Even when prior experience is taken into account, participants still require orientation to the specific interaction context of the study. Structured onboarding tasks that mirror the main study activities help participants understand how to interact with the system and adapt existing habits to new modalities. When onboarding aligns with task goals---such as editing images or navigating maps---participants are more willing to experiment with unfamiliar input methods. In \textbf{Case D}, a warm-up task that incorporated text, scribbles, and annotations closely mirrored the main study and helped participants use a broader range of inputs later. By contrast, in \textbf{Case B}, a scripted onboarding that focused on a generic voice example did not sufficiently prepare participants to use the GUI-referencing feature, leading them to rely on familiar command-style interactions. Aligning onboarding more closely with actual study tasks can therefore help participants bridge the gap between prior experience and the intended interaction design, promoting richer exploration and engagement.

\paragraph{R1.4 Offer a low-pressure trial phase before formal tasks}
After completing structured onboarding, participants often still need space to explore new input methods without performance pressure. A short \textbf{exploratory trial phase} before formal tasks allows participants to internalize what they learned during onboarding and to build confidence in how the system interprets their actions. This phase can include unstructured interaction as well as light demonstrations, such as showing how a scribble is recognized or presenting example prompts. In \textbf{Case B}, participants who lacked early opportunities for open-ended exploration worried about providing ``incorrect'' input, whereas in \textbf{Case D}, those who were able to experiment freely beforehand became more confident using scribbles and annotations during the main tasks. Providing a low-pressure exploration phase helps participants translate structured learning into flexible engagement, reducing anxiety and fostering more authentic interaction with unfamiliar inputs.

\subsection{G2. Align Prototype Fidelity to Study Goals}  
Balancing prototype fidelity with experimental control is a recurring challenge in evaluating GenAI systems (see C2, \autoref{sec:c2}). High-fidelity prototypes enable authentic interactions but introduce unpredictable outputs that compromise consistency, whereas low-fidelity setups increase control but sacrifice the generative qualities that define GenAI. This trade-off, therefore, becomes a methodological choice rather than a purely technical one.
The five recommendations for G2 outline how to align fidelity with study goals, address system variability, and maintain both experimental rigor and ecological validity.

\paragraph{R2.1 Choose prototype fidelity according to study objectives.}
Prototype fidelity should align with the main research goal, whether to explore input preferences, observe adaptation to system behavior, or evaluate output quality. In early-stage investigations that focus on \textbf{input exploration} or interaction patterns, simplified or non-functional prototypes help isolate user strategies. This avoids confounding effects such as latency or unpredictable output.
In \textbf{Case C}, a paper-based prototype was used to examine input choices (text, scribbles, annotations) without system interference, enabling clearer observation of interaction patterns.
In contrast, in \textbf{Case D}, a fully functional image generation prototype allowed researchers to observe real-time adaptation but also revealed issues with latency and output errors. 
Choosing fidelity with intention helps ensure methodological coherence between research goals, study conditions, and interpretive validity.

\paragraph{R2.2 Manage unpredictability through selective backend adjustments.}
Functional prototypes are essential when studying how users adapt to GenAI systems in realistic conditions, such as refining prompts or responding to variable outputs. Although they reduce experimental control, they reveal interaction patterns that only emerge during live system use. To balance realism and reliability, researchers can manage unpredictability through selective adjustments to the backend, such as refining prompts, tuning model parameters, or adding contextual guidance. 
In \textbf{Case D}, a fully functional prototype supported real-time image generation from text, scribbles, and annotations. To reduce excessive output variability while preserving generative behavior, backend prompts were constrained during closed-ended tasks to better align with task goals, whereas fewer prompt constraints were applied during open-ended phases to allow broader exploration. This setup enabled observation of how participants refined their inputs in response to varied outputs under different degrees of system control. In \textbf{Case A}, backend adjustments to the voice assistant prompt (e.g., including navigation examples) reduced hallucinations without diminishing the system's perceived authenticity.
Balancing fidelity and control through targeted technical adjustments helps researchers capture GenAI-specific interaction dynamics while maintaining interpretive validity across participants. 

\paragraph{R2.3 Prepare fallback strategies proactively to sustain study flow.}
\label{sec:r2.3}
Even with backend control, GenAI systems can fail to produce coherent or timely outputs. Such disruptions can interrupt task flow and frustrate participants, thereby reducing the reliability of collected data. We therefore recommend designing \textbf{fallback strategies}, such as pre-generated outputs, scripted alternatives, or structured opportunities to re-prompt, to maintain task continuity when system breakdowns occur. In \textbf{Case D}, participants were allowed to re-prompt the system when image generation failed, which helped them remain engaged and complete the task despite interruptions. During analysis, the research team noted that preparing additional fallback materials, such as pre-generated images, could further support continuity in similar studies. By contrast, in \textbf{Case B}, the absence of fallback options forced participants to restart interactions manually, leading to frustration and loss of focus. Overall, proactive fallback planning supports both data quality and participant engagement, while documenting such events enables richer post-study analysis of system breakdowns and recovery strategies.

\paragraph{R2.4 Document system behavior and contextual variables.}
Because GenAI systems can produce variable and unpredictable responses, detailed documentation of system behavior is essential for credible interpretation and replication. Logging prompts, outputs, latency, and contextual variables allows researchers to understand how system performance shapes participants' experiences and to attribute observed behaviors more accurately. In \textbf{Case D}, extensive event logging and synchronized voice recordings enabled the research team to trace each interaction and examine how users adapted to system responses in real time. This comprehensive documentation provided valuable insights during analysis and helped differentiate between user behavior, interface design, and stochastic model variation. Overall, thorough and transparent logging ensures that GenAI evaluation remains interpretable and reproducible, enabling researchers to distinguish between design-related issues and variability inherent to generative systems.

\paragraph{R2.5 Design tasks that reflect GenAI's exploratory nature.}
Study tasks should represent how users naturally engage with GenAI systems through iteration, experimentation, and adaptation. Restricting interaction to single attempts limits realism and prevents observation of how participants refine their input in response to system outputs.
In \textbf{Case D}, participants frequently modified prompts or sketches in response to system feedback but were constrained by a study-imposed three-iteration limit, which shaped their interaction strategies. By contrast, \textbf{Case B} deliberately incorporated an open-ended task that allowed participants to freely ask questions to the in-car assistant, revealing more spontaneous exploration patterns. Similarly, in \textbf{Case A}, participants interacted with the system through loosely defined goals, such as setting destinations, asking about car functions, or initiating casual conversation. To support hesitant participants, the research team prepared a small set of fallback ideas to help maintain engagement. Designing tasks that support structured yet flexible iteration better captures the exploratory nature of GenAI use and provides more ecologically valid insights into user adaptation. Choosing appropriate prototype fidelity, managing unpredictability transparently, and enabling iterative exploration together help preserve both experimental rigor and the authentic dynamics of generative interaction.

\subsection{G3. Improve Feedback Interpretability and User Trust}  
As identified in Challenge C3 (\autoref{sec:c3}), GenAI systems \textbf{redefine} feedback. Their responses are not always consistent or interpretable, which can make it difficult for participants to form stable mental models or trust the system's behavior. Unlike conventional interfaces, where the relationship between input and output is transparent, GenAI feedback can be ambiguous, delayed, or seemingly unrelated to the input. The three recommendations under G3 address feedback interpretability as a continuous process: clarifying how the feedback loop operates across input and output (R3.1), providing real-time cues that confirm input recognition across modalities (R3.2), and calibrating participant expectations regarding feedback reliability (R3.3).

\paragraph{R3.1 Make system feedback loop interpretable across input and output.}
Participants often struggle to understand whether their input was received and why outputs vary, particularly when feedback is delayed or inconsistent. Establishing transparency throughout the feedback loop---by acknowledging inputs and contextualizing output variation---can help participants build trust and maintain engagement. Providing clear indicators of input recognition (e.g., progress cues or acknowledgment tones), along with brief explanations of why outputs differ, can support a more predictable interaction flow. 
In \textbf{Case B}, latency and missing confirmation cues led some participants to repeat voice commands, unsure whether their input had been processed. In \textbf{Case D}, participants initially viewed inconsistent image results as system errors; however, brief clarifications that the model was intentionally exploring multiple interpretations helped participants understand output variation as a design feature rather than as an error. In \textbf{Case A}, subtle acknowledgments of recognized speech and contextually adaptive phrasing appeared to encourage participants to view response differences as flexibility rather than faults. Across these cases, these examples suggest that making feedback interpretable across both input and output can reduce uncertainty and support user confidence.

\paragraph{R3.2 Provide real-time input feedback for immediate transparency.}
Moment-to-moment uncertainty often arises when participants are unsure whether the system has captured their input or is still processing it. This hesitation can disrupt task flow, particularly in modalities such as voice or sketches where input recognition is less visible.
We recommend implementing \textbf{real-time feedback cues} that visualize how the system interprets user input, such as speech-to-text transcriptions, highlighted drawings, or short textual summaries of recognized content. Such cues can provide immediate reassurance that input has been processed and allow participants to stay focused on evaluating system behavior. In \textbf{Cases A and B}, on-screen transcriptions improved transparency but did not fully prevent repeated queries during latency delays, as users sometimes remained unsure when the assistant was ``listening.'' In \textbf{Case D}, participants tended to trust text prompts more than scribbles because textual input offered visible acknowledgment, while sketches lacked explicit confirmation. Providing timely, modality-specific feedback can therefore help minimize hesitation and support smoother, more confident engagement.

\paragraph{R3.3 Use post-task debriefs to identify mismatches between user intent and system behavior.}
Not all misunderstandings between participants and GenAI systems are visible during interaction. Participants may misinterpret responses, question their own input, or blame themselves without expressing this uncertainty in real-time. Structured post-task debriefs can help identify and surface hidden mismatches between user intent and system behavior. We recommend a brief set of follow-up questions after each task, such as ``What did you expect the system to do?'' or ``What do you think the system understood?'' These reflections can reveal unspoken confusion and help researchers interpret observed behaviors more accurately. In \textbf{Case D}, participants sometimes attributed unexpected images to their ``bad writing,'' later explaining that they were unsure whether the input had been recognized. Similarly, in \textbf{Case B}, post-task interviews revealed uncertainty about whether spoken inputs were understood when responses were delayed or off-topic. Post-task debriefs thus provide important context for interpreting user behavior beyond what is directly observable during the study.

\subsection{G4. Adapt Evaluation Strategies to Capture GenAI-Specific User Experiences} 
As discussed in Challenge C4 (\autoref{sec:c4}), evaluating GenAI systems using traditional usability metrics, such as SUS or UEQ, can lead to incomplete or potentially misleading conclusions. Because GenAI outputs are variable and sometimes hallucinated, user frustration or confusion may stem from model behavior rather than from interface design alone. Conventional usability scales assume consistent, deterministic system responses, which can limit their ability to capture GenAI-specific phenomena such as unpredictability, trust, or intent alignment. The two recommendations under G4 outline how to adapt evaluation strategies to these conditions: by extending what is measured with GenAI-specific constructs (R4.1) and by combining standardized metrics with qualitative reflections for deeper interpretability (R4.2).

\paragraph{R4.1 Expand evaluation metrics to capture GenAI-specific constructs.}
Traditional usability scales quantify satisfaction and efficiency but often overlook experiential factors central to GenAI interaction, such as trust, confidence, intent alignment, and comfort with uncertainty. We therefore recommend integrating targeted questions such as ``Did you trust the system’s output?'', ``How confident were you that your input was understood?'', or ``Did the result match your intent?'' to better reflect these GenAI-specific aspects of user experience. In \textbf{Case B}, post-task reflections indicated that participants' satisfaction was driven less by response accuracy and more by how ``understood'' they felt by the voice assistant. Similarly, in \textbf{Case D}, participants sometimes rated the same image output differently depending on whether they believed the system had captured their intent. Including items that capture perceived understanding or intent alignment (e.g., ``I felt the system understood what I meant'') can provide quantifiable yet context-specific data that complements standard usability metrics.

\paragraph{R4.2 Pair standardized metrics with qualitative reflections.}
While SUS and UEQ summarize user perceptions quantitatively, they rarely reveal whether ratings primarily reflect interface design or GenAI-specific variability. We recommend pairing standardized scores with short, open-ended reflections after each task, complemented by observation or think-aloud protocols. Follow-up prompts such as ``What did you expect to happen?'' or ``Was the response what you intended?'' can help clarify how participants interpreted their experiences. In \textbf{Cases A and B}, SUS and UEQ offered a general overview of usability, but interviews suggested that lower scores were often influenced by latency or hallucinations rather than by input design. In \textbf{Cases C and D}, think-aloud sessions revealed uncertainty about whether scribbles were interpreted, even in the absence of formal metrics. Combining structured ratings with qualitative insights thus helps ensure that findings reflect both the measurable and interpretive aspects of GenAI interactions.

\subsection{G5. Build Flexibility into GenAI Study Design and Analysis}
As discussed in Challenge C5 (\autoref{sec:c5}), distinguishing usability issues from GenAI-specific limitations, such as hallucinations, latency, or backend instability, proved difficult in our studies. These ambiguities complicate analysis and call for flexible, reflexive study designs. Researchers, therefore, need to be prepared to adapt tasks, logging, or analytic strategies in response to unexpected system behavior or participant confusion. The three recommendations under G5 highlight how flexibility can be integrated throughout a study: proactively monitoring system reliability (R5.1), maintaining task flow during disruptions (R5.2), and labeling the system's limitations to support accurate interpretation (R5.3).

\paragraph{R5.1 Anticipate system issues through pilot testing and adapt during live monitoring.}
GenAI systems can fail unpredictably, producing long delays, repeated hallucinations, or instability that disrupts the task flow. We recommend \textbf{monitoring system behavior continuously} during both pilot testing and live sessions to help detect issues early and enable timely adjustments. This can include logging outputs in real-time and preparing adaptive responses, such as prompt modifications or fallback content.
In \textbf{Case B}, pilot feedback identified latency as a major issue, and iterative adjustments helped reduce delays that might otherwise have been mistaken for usability flaws. In \textbf{Case D}, pre-study testing revealed performance differences between two image-generation models, and the team alternated between the two models to maintain acceptable responsiveness while preserving realistic interaction behavior. Continuous monitoring and real-time intervention can help ensure that technical failures do not unduly distort user evaluation.

\paragraph{R5.2 Respond flexibly to system failures to preserve study continuity.}
Unlike R2.3 (\autoref{sec:r2.3}), which emphasizes proactively preparing fallback strategies before a study to reduce unpredictability, this recommendation focuses on \textbf{reactive adaptation during data collection}. When GenAI systems fail or generate unusable outputs, continuing the original task can frustrate participants and compromise data quality. Researchers may therefore need to adjust tasks, repeat inputs, or offer workarounds to sustain engagement and support meaningful data collection despite disruptions. We recommend preparing \textbf{alternative task paths}, such as allowing retries, providing pre-generated outputs, or skipping tasks when failures persist, as well as \textbf{time-boxing activities} to balance flexibility with session duration. All in-session adaptations should be \textbf{documented} so they can be considered during later analysis.
In \textbf{Case D}, participants could retry image generation up to three times; if failures persisted, they were guided to skip the task. This approach helped maintain flow and later prompted discussion about the potential value of pre-generated examples for continuity. In \textbf{Case B}, participants were allowed to retry tasks or restart voice input using the push-to-talk button, providing a simple yet effective recovery mechanism. Such reactive flexibility can prevent technical breakdowns from derailing studies and help ensure that adaptive decisions are captured and reflected during subsequent interpretation.

\paragraph{R5.3 Label system limitations in logs to ensure transparent analysis.}
To interpret study results accurately, researchers need to separate interface-related challenges from system-side issues. We recommend systematically labeling known limitations, such as ``latency $>$ 3s,'' ``output failures,'' or ``hallucination detected,'' within session logs. Such annotations help clarify when participant hesitation or performance drops are attributable to system behavior rather than interface design.
In \textbf{Case D}, logging which model generated each image helped differentiate latency-related pauses from genuine interaction difficulties. In \textbf{Case B}, separating sessions with and without hallucinations clarified which usability scores reflected user experience versus technical artifacts. These annotations create an ``audit trail'' for later analysis, supporting more reliable interpretation and facilitating replication by other researchers. Labeling the system's limitations in this way can enhance analytic transparency and strengthen confidence in the reported findings.

\section{Discussion}
Our findings highlight challenges and recommendations that illustrate how GenAI systems are reshaping controlled lab studies and HCI evaluation. Below, we discuss how these shifts affect study design, user trust, and evaluation metrics, and we reflect on how a reflective multi-case approach helps surface emerging methodological challenges.

\subsection{How GenAI Evaluation Extends and Reframes Prior HCI Evaluation}
Our results extend long-standing HCI debates about fidelity, control, and trust. As in prior work on adaptive or novel-sensing systems that require context-sensitive and longitudinal observation~\cite{poppe_evaluating_2007, rogers_case_2017, kjaerup_longitudinal_2021}, GenAI variability disrupts the tight input--output mappings assumed by many lab protocols. It also intensifies the fidelity--control trade-offs familiar from Wizard-of-Oz and prototyping research~\cite{klemmer_suede_2000}. GenAI does not merely introduce non-determinism that complicates study design; it \emph{can} also generate new challenges while amplifying and reframing existing ones.

Several tensions identified in our analysis---such as users reverting to familiar interaction strategies, difficulties interpreting system behavior, and trade-offs between control and realism---have also been discussed in evaluations of other adaptive or intelligent systems, including speech recognition and recommendation systems~\cite{buschek_how_2022, chen_screen_2023, storms_transparency_2022}. Our contribution is to articulate how the high degree of output variability in GenAI systems amplifies these tensions by coupling probabilistic generation with feedback ambiguity. For example, variability may support exploration in some contexts while undermining trust in others, including through hallucinations. We therefore frame these challenges as amplified or reframed methodological concerns, or issues that emerge specifically in the context of generative systems, rather than as entirely unprecedented problems.
Challenges C1--C3 primarily stem from stochastic outputs and the lack of clear input--output mappings. Unpredictable feedback can discourage deviation from familiar strategies (C1) and disrupt fidelity--control trade-offs (C2) and feedback loops (C3), thereby undermining both user confidence and experimental control. Challenges C4--C5 arise from the opacity of system behavior and the entanglement of model and interface effects, complicating evaluation and prompting a reconsideration of how success and usability are defined. 

One could argue that non-determinism ultimately reflects training data and tunable randomness (e.g., temperature or sampling strategies). However, we contend that this unpredictability is precisely what makes these systems worth studying: variability enables open-ended dialogue, exploration of alternatives, and forms of collaboration that deterministic systems cannot support. Rather than suppressing variability, researchers should examine whether systems support exploration and recovery, and how users adapt interaction patterns over multiple steps to reach their goals.
In considering new approaches to studying GenAI, we argue that methodological transparency is critical. We examined our own research process---including the decisions made, trade-offs encountered, and evolving strategies---and how these choices shaped study procedures and outcomes. Decisions such as constraining inputs, simulating outputs, or tolerating variability were not neutral in their effects; they influenced what could be observed and how findings were interpreted. We therefore recommend reflective case comparisons as a useful way to surface methodological insights for emerging technologies. Looking ahead, as models expand their context windows and adopt more agentic behaviors, additional challenges will likely require longitudinal and in-the-wild studies to understand how trust, adaptation, and success criteria evolve over time.

\subsection{Study Design Strategies for User Trust and Confidence}
Even when the underlying GenAI model is opaque, a well-designed study can help foster transparency for participants. Across our case studies, three strategies proved particularly effective. First, onboarding that emphasized input possibilities rather than the system's ``AI'' identity encouraged exploration and reduced hesitation when trying unfamiliar input methods. Second, during interaction, real-time feedback cues---such as transcribed speech or visual confirmations of recognized input---helped participants understand how their actions were interpreted and reassured them that their input had been received. These strategies were especially important when participants encountered unexpected outputs that stemmed not from their input or interface design, but from the system's generative behavior. Making this distinction explicit during the study helped reduce confusion and provided a clearer basis for interpreting participants' responses. Finally, structured post-task reflection questions revealed mismatches between participants' intentions and the system's responses that were not apparent from interaction logs alone. These strategies illustrate how transparency can be purposefully designed into GenAI evaluations, helping participants navigate uncertainty while enabling researchers to interpret behavior more reliably.

\subsection{Rethinking What and How We Measure in GenAI Evaluation} 
Conventional usability scales (e.g., SUS, UEQ) remain useful for comparability, but in our GenAI studies, they often lacked expressiveness for the underlying issues that users encountered. Low usability scores could result from interface design, unpredictable outputs, hallucinations, or mismatches between user intentions and system responses. Relying solely on these standard metrics may therefore fail to capture the nuanced sources of user frustration and can lead to misinterpretation. To address this limitation, we suggest extending standard scales with GenAI-specific constructs such as user trust, confidence, and intent alignment. Adding short probes after each task (e.g., ``What did you expect to happen?'' or ``How do you think the system understood your input?'') helped contextualize numerical ratings and revealed mismatches that were otherwise invisible in logs or scale data. Our studies further showed that different methods illuminate different layers of interaction. Usability ratings summarized perceptions of ease, efficiency, and satisfaction; interviews explained participants’ reasoning behind those scores; and observations highlighted challenges that emerged during real interactions. While each method contributes a distinct perspective, relying on a single approach can lead to blind spots. This aligns with long-standing calls in HCI and the social sciences for methodological triangulation, in which multiple methods are combined to compensate for individual limitations and build a more complete picture of interaction~\cite{mcgrath_methodology_1995, van_turnhout_design_2014}. For GenAI evaluation, mixed-method strategies, such as combining metrics with reflections or triangulating observation, can support more reliable interpretation and help distinguish interface-related issues from GenAI-specific effects~\cite{yin_understanding_2019}.

\subsection{Limitations and Future Work}
Our findings are derived from four short-term, lab-based studies with specific prototypes and user groups. This scope allowed us to identify recurring methodological challenges, but it constrains the empirical generalizability of our findings and the techno-ecological validity of our conclusions. Accordingly, our contribution should be understood as methodological insights derived through cross-case reasoning, rather than as claims that are empirically replicable or statistically generalizable. Nonetheless, highlighting these patterns within constrained settings offers a valuable starting point for understanding how GenAI complicates established lab-based evaluation practices. Additionally, GenAI systems and users' mental models evolve rapidly. Some challenges may diminish as model capabilities improve or as users become more proficient with GenAI tools. We did not study long-term adaptation, in-the-wild use, or a broader range of application domains, all of which represent important directions for future work and motivate extending analyses to a more diverse set of case studies. Such extensions may surface additional challenges or complement those reported here. We therefore do not aim to provide an exhaustive review but instead report a reflective cross-case analysis grounded in an in-depth understanding of the research process. To make our analysis steps and underlying reflections transparent, we provide the coding schema and theme descriptions in the supplementary materials. While these materials document how the analysis was conducted, reproducing the cross-case analysis depends on access to a similarly structured study process. For this reason, we encourage documenting methodological trade-offs and reflections in a comparable manner.

Future studies could explore several additional avenues. One potential direction is to conduct comparative user studies of different evaluation strategies for GenAI systems. Variables such as fallback strategies (retrying versus pre-generated outputs), onboarding framings (AI-highlighted versus task-highlighted instructions), and feedback cues (verbatim transcripts versus interpreted summaries) may influence how participants develop trust, confidence, and input strategies. Such comparisons can help clarify when particular study designs lead to distinct user behaviors and refine best practices in GenAI evaluation. 
Another direction is to explore GenAI not only as a system under evaluation, but also as a research tool within HCI studies. Recent work has examined the use of LLMs as simulated participants~\cite{kapania_simulacrum_2025} and as aids in qualitative analysis, such as thematic analysis~\cite{dai_llm---loop_2023}. Other studies have compared how human analysts classify qualitative data with how LLMs generate classifications and reasoning for the same material~\cite{bano_exploring_2023}. These efforts raise additional methodological questions about validity, interpretation, and researcher responsibility, suggesting that reflective evaluation practices will remain important as GenAI becomes increasingly embedded in the research process itself. Finally, we view our contribution as an initial step rather than a definitive account. Additional challenges and recommendations are likely to emerge as GenAI systems become more prevalent in everyday contexts, underscoring the need to revisit and refine evaluation practices over time.

\section{Conclusion}
In this work, we identified five recurring methodological challenges (C1--C5) in evaluating GenAI systems in controlled lab settings through a reflective analysis of four case studies. These challenges reveal how GenAI's stochastic, non-deterministic behavior complicates established assumptions in HCI evaluation. In doing so, they expose tensions between control and realism, between deterministic expectations and probabilistic system behavior, and between user interpretation and model variability.
Building on these insights, we proposed five methodological guidelines (G1--G5) and eighteen practice-oriented recommendations to support researchers in designing, conducting, and analyzing GenAI user studies more effectively. These guidelines include preparing participants for unpredictable behavior, aligning prototype fidelity with study goals, improving feedback interpretability, adapting evaluation metrics to account for stochasticity, and incorporating flexibility and transparency into study design and analysis.
Overall, our work foregrounds the research process itself---its design choices, trade-offs, and interpretive challenges---when studying generative systems. Rather than offering a final framework, we aim to provide a foundation for continued methodological reflection as GenAI technologies become more integrated into everyday life. Continually revisiting and expanding these methodological discussions will be essential for building reliable and transparent HCI research on GenAI-integrated systems.

\begin{acks}
The authors used generative AI tools to improve readability (e.g., spelling, rephrasing, and formatting). All substantive content, analysis, and conclusions were developed by the authors, who retain full responsibility for the work.
\end{acks}

\bibliographystyle{ACM-Reference-Format}
\bibliography{ref}

@article{almulla_understanding_2025,
	title        = {Understanding the Challenges and Promises of Developing Generative AI Apps: An Empirical Study},
	author       = {AlMulla, Buthayna and Assi, Maram and Hassan, Safwat},
	year         = {2025},
	journal      = {arXiv preprint arXiv:2506.16453},
	doi          = {10.48550/arXiv.2506.16453},
	url          = {http://arxiv.org/abs/2506.16453}
}

@article{bano_exploring_2023,
	title        = {Exploring qualitative research using LLMs},
	author       = {Bano, Muneera and Zowghi, Didar and Whittle, Jon},
	year         = {2023},
	journal      = {arXiv preprint arXiv:2306.13298},
	doi          = {10.48550/arXiv.2306.13298},
	url          = {http://arxiv.org/abs/2306.13298}
}

@article{bertini_beliv08_2007,
	title        = {BELIV'06: beyond time and errors; novel evaluation methods for information visualization},
	author       = {Bertini, Enrico and Plaisant, Catherine and Santucci, Giuseppe},
	year         = {2007},
	journal      = {Interactions},
	publisher    = {ACM New York, NY, USA},
	volume       = {14},
	number       = {3},
	pages        = {59--60},
	doi          = {10.1145/1358628.1358955},
	url          = {https://dl.acm.org/doi/10.1145/1358628.1358955}
}

@article{beyer_contextual_1999,
	title        = {Contextual design},
	author       = {Beyer, Hugh and Holtzblatt, Karen},
	year         = {1999},
	month        = jan,
	journal      = {Interactions},
	publisher    = {Association for Computing Machinery},
	address      = {New York, NY, USA},
	volume       = {6},
	number       = {1},
	pages        = {32--42},
	doi          = {10.1145/291224.291229},
	issn         = {1072-5520},
	url          = {https://doi.org/10.1145/291224.291229},
	issue_date   = {Jan./Feb. 1999},
	numpages     = {11}
}

@book{blandford_qualitative_2016,
	title        = {Qualitative HCI research: Going behind the scenes},
	author       = {Blandford, Ann and Furniss, Dominic and Makri, Stephann},
	year         = {2016},
	publisher    = {Morgan \& Claypool Publishers},
	isbn         = {978-1-62705-760-8}
}

@article{braun_reflecting_2019,
	title        = {Reflecting on reflexive thematic analysis},
	author       = {Braun, Virginia and Clarke, Victoria},
	year         = {2019},
	journal      = {Qualitative research in sport, exercise and health},
	publisher    = {Taylor \& Francis},
	volume       = {11},
	number       = {4},
	pages        = {589--597},
	doi          = {10.1080/2159676X.2019.1628806},
	url          = {https://www.tandfonline.com/doi/full/10.1080/2159676X.2019.1628806}
}

@article{braun_using_2006,
	title        = {Using thematic analysis in psychology},
	author       = {Braun, Virginia and Clarke, Victoria},
	year         = {2006},
	journal      = {Qualitative research in psychology},
	publisher    = {Taylor \& Francis},
	volume       = {3},
	number       = {2},
	pages        = {77--101},
	doi          = {10.1191/1478088706qp063oa},
	issn         = {1478-0887, 1478-0895},
	url          = {http://www.tandfonline.com/doi/abs/10.1191/1478088706qp063oa}
}

@article{brdnik_intelligent_2022,
	title        = {Intelligent user interfaces and their evaluation: a systematic mapping study},
	author       = {Brdnik, Sa{\v{s}}a and Heri{\v{c}}ko, Tja{\v{s}}a and {\v{S}}umak, Bo{\v{s}}tjan},
	year         = {2022},
	journal      = {Sensors},
	publisher    = {MDPI},
	volume       = {22},
	number       = {15},
	pages        = {5830},
	doi          = {10.3390/s22155830},
	url          = {https://www.mdpi.com/1424-8220/22/15/5830}
}

@inproceedings{brown_language_2020,
	title        = {Language {Models} are {Few}-{Shot} {Learners}},
	author       = {Brown, Tom and Mann, Benjamin and Ryder, Nick and Subbiah, Melanie and Kaplan, Jared D and Dhariwal, Prafulla and Neelakantan, Arvind and Shyam, Pranav and Sastry, Girish and Askell, Amanda and Agarwal, Sandhini and Herbert-Voss, Ariel and Krueger, Gretchen and Henighan, Tom and Child, Rewon and Ramesh, Aditya and Ziegler, Daniel and Wu, Jeffrey and Winter, Clemens and Hesse, Chris and Chen, Mark and Sigler, Eric and Litwin, Mateusz and Gray, Scott and Chess, Benjamin and Clark, Jack and Berner, Christopher and McCandlish, Sam and Radford, Alec and Sutskever, Ilya and Amodei, Dario},
	year         = 2020,
	booktitle    = {Advances in {Neural} {Information} {Processing} {Systems}},
	publisher    = {Curran Associates, Inc.},
	volume       = 33,
	pages        = {1877--1901},
	url          = {https://papers.nips.cc/paper/2020/hash/1457c0d6bfcb4967418bfb8ac142f64a-Abstract.html},
	urldate      = {2026-01-21}
}

@article{buschek_how_2022,
	title        = {How to Support Users in Understanding Intelligent Systems? An Analysis and Conceptual Framework of User Questions Considering User Mindsets, Involvement, and Knowledge Outcomes},
	author       = {Buschek, Daniel and Eiband, Malin and Hussmann, Heinrich},
	year         = {2022},
	month        = nov,
	journal      = {ACM Trans. Interact. Intell. Syst.},
	publisher    = {Association for Computing Machinery},
	address      = {New York, NY, USA},
	volume       = {12},
	number       = {4},
	doi          = {10.1145/3519264},
	issn         = {2160-6455},
	url          = {https://doi.org/10.1145/3519264},
	issue_date   = {December 2022},
	articleno    = {29},
	numpages     = {27}
}

@inproceedings{capdevila_current_2021,
	title        = {Do current user testing practices meet the needs of the new interactive paradigms?},
	author       = {Capdevila, Marc Gonzalez and Saltiveri, Toni Granollers and Garrido, Juan Enrique and M\"{u}ller, Oct\'{a}vio Henrique and Ruas, Leonardo Coelho},
	year         = {2021},
	booktitle    = {Proceedings of the XXI International Conference on Human Computer Interaction},
	location     = {M\'{a}laga, Spain},
	publisher    = {Association for Computing Machinery},
	address      = {New York, NY, USA},
	series       = {Interacci\'{o}n '21},
	doi          = {10.1145/3471391.3471416},
	isbn         = {9781450375979},
	url          = {https://doi.org/10.1145/3471391.3471416},
	articleno    = {22},
	numpages     = {9}
}

@inproceedings{chen_screen_2023,
	title        = {Screen or No Screen? Lessons Learnt from a Real-World Deployment Study of Using Voice Assistants With and Without Touchscreen for Older Adults},
	author       = {Chen, Chen and Lifset, Ella T and Han, Yichen and Roy, Arkajyoti and Hogarth, Michael and Moore, Alison A and Farcas, Emilia and Weibel, Nadir},
	year         = {2023},
	booktitle    = {Proceedings of the 25th International ACM SIGACCESS Conference on Computers and Accessibility},
	location     = {New York, NY, USA},
	publisher    = {Association for Computing Machinery},
	address      = {New York, NY, USA},
	series       = {ASSETS '23},
	doi          = {10.1145/3597638.3608378},
	isbn         = {9798400702204},
	url          = {https://doi.org/10.1145/3597638.3608378},
	articleno    = {52},
	numpages     = {21}
}

@incollection{cooper_thematic_2012,
	title        = {Thematic analysis.},
	author       = {Braun, Virginia and Clarke, Victoria},
	year         = {2012},
	booktitle    = {{APA} handbook of research methods in psychology, {Vol} 2: {Research} designs: {Quantitative}, qualitative, neuropsychological, and biological.},
	publisher    = {American Psychological Association},
	address      = {Washington},
	pages        = {57--71},
	doi          = {10.1037/13620-004},
	isbn         = {978-1-4338-1005-3},
	url          = {https://content.apa.org/books/13620-004},
	urldate      = {2024-12-20},
	language     = {en},
	editor       = {Cooper, Harris and Camic, Paul M. and Long, Debra L. and Panter, A. T. and Rindskopf, David and Sher, Kenneth J.}
}

@book{creswell_designing_2017,
	title        = {Designing and conducting mixed methods research},
	author       = {Creswell, John W and Clark, Vicki L Plano},
	year         = {2017},
	publisher    = {Sage publications}
}

@article{dahlback_wizard_1993,
	title        = {Wizard of Oz studies — why and how},
	author       = {Dahlb\"{a}ck, N. and J\"{o}nsson, A. and Ahrenberg, L.},
	year         = {1993},
	month        = dec,
	journal      = {Know.-Based Syst.},
	publisher    = {Elsevier Science Publishers B. V.},
	address      = {NLD},
	volume       = {6},
	number       = {4},
	pages        = {258–266},
	doi          = {10.1016/0950-7051(93)90017-N},
	issn         = {0950-7051},
	url          = {https://doi.org/10.1016/0950-7051(93)90017-N},
	issue_date   = {December 1993},
	numpages     = {9}
}

@article{dai_llm---loop_2023,
	title        = {LLM-in-the-loop: Leveraging large language model for thematic analysis},
	author       = {Dai, Shih-Chieh and Xiong, Aiping and Ku, Lun-Wei},
	year         = {2023},
	journal      = {arXiv preprint arXiv:2310.15100},
	doi          = {10.48550/arXiv.2310.15100},
	url          = {https://arxiv.org/abs/2310.15100}
}

@inproceedings{dalsgaard_reflective_2012,
	title        = {Reflective design documentation},
	author       = {Dalsgaard, Peter and Halskov, Kim},
	year         = {2012},
	booktitle    = {Proceedings of the Designing Interactive Systems Conference},
	location     = {Newcastle Upon Tyne, United Kingdom},
	publisher    = {Association for Computing Machinery},
	address      = {New York, NY, USA},
	series       = {DIS '12},
	pages        = {428–437},
	doi          = {10.1145/2317956.2318020},
	isbn         = {9781450312103},
	url          = {https://doi.org/10.1145/2317956.2318020},
	numpages     = {10}
}

@article{eisenhardt_building_1989,
	title        = {Building theories from case study research},
	author       = {Eisenhardt, Kathleen M},
	year         = {1989},
	journal      = {Academy of management review},
	publisher    = {Academy of Management Briarcliff Manor, NY 10510},
	volume       = {14},
	number       = {4},
	pages        = {532--550},
	doi          = {10.2307/258557},
	url          = {https://www.jstor.org/stable/258557}
}

@incollection{fulfagar_development_2021,
	title        = {Development and Evaluation of Usability Heuristics for Voice User Interfaces},
	author       = {Fulfagar, Lokesh and Gupta, Anupriya and Mathur, Arpit and Shrivastava, Abhishek},
	year         = {2021},
	booktitle    = {Design for Tomorrow -- Volume 1},
	publisher    = {Springer},
	series       = {Smart Innovation, Systems and Technologies},
	volume       = {221},
	pages        = {375--385},
	doi          = {10.1007/978-981-16-0041-8_32},
	url          = {https://doi.org/10.1007/978-981-16-0041-8_32}
}

@article{gmeiner_prototyping_2025,
	title        = {Prototyping Multimodal GenAI Real-Time Agents with Counterfactual Replays and Hybrid Wizard-of-Oz},
	author       = {Gmeiner, Frederic and Holstein, Kenneth and Martelaro, Nikolas},
	year         = {2025},
	journal      = {arXiv preprint arXiv:2510.06872},
	doi          = {10.48550/arXiv.2510.06872},
	url          = {https://arxiv.org/abs/2510.06872}
}

@inproceedings{greenberg_usability_2008,
	title        = {Usability evaluation considered harmful (some of the time)},
	author       = {Greenberg, Saul and Buxton, Bill},
	year         = {2008},
	booktitle    = {Proceedings of the SIGCHI Conference on Human Factors in Computing Systems},
	location     = {Florence, Italy},
	publisher    = {Association for Computing Machinery},
	address      = {New York, NY, USA},
	series       = {CHI '08},
	pages        = {111–120},
	doi          = {10.1145/1357054.1357074},
	isbn         = {9781605580111},
	url          = {https://doi.org/10.1145/1357054.1357074},
	numpages     = {10}
}

@misc{gustafsson_single_2017,
	title        = {Single Case Studies vs. Multiple Case Studies: A Comparative Study},
	author       = {Gustafsson, Johanna},
	year         = 2017,
	note         = {Literature review, Academy of Business, Engineering and Science, Halmstad University, Sweden}
}

@inproceedings{he_ai_2024,
	title        = {AI and the Future of Collaborative Work: Group Ideation with an LLM in a Virtual Canvas},
	author       = {He, Jessica and Houde, Stephanie and Gonzalez, Gabriel E. and Silva Moran, Dar\'{\i}o Andr\'{e}s and Ross, Steven I. and Muller, Michael and Weisz, Justin D.},
	year         = {2024},
	booktitle    = {Proceedings of the 3rd Annual Meeting of the Symposium on Human-Computer Interaction for Work},
	location     = {Newcastle upon Tyne, United Kingdom},
	publisher    = {Association for Computing Machinery},
	address      = {New York, NY, USA},
	series       = {CHIWORK '24},
	doi          = {10.1145/3663384.3663398},
	isbn         = {9798400710179},
	url          = {https://doi.org/10.1145/3663384.3663398},
	articleno    = {9},
	numpages     = {14}
}

@article{iniguez-carrillo_usability_2021,
	title        = {Usability questionnaires to evaluate voice user interfaces},
	author       = {Iniguez-Carrillo, Adriana Lorena and Gaytan-Lugo, Laura Sanely and Garcia-Ruiz, Miguel Angel and Maciel-Arellano, Rocio},
	year         = {2021},
	journal      = {IEEE Latin America Transactions},
	publisher    = {IEEE},
	volume       = {19},
	number       = {9},
	pages        = {1468--1477},
	doi          = {10.1109/TLA.2021.9468439}
}

@inproceedings{jansen_wizard_2022,
	title        = {Wizard of Errors: Introducing and Evaluating Machine Learning Errors in Wizard of Oz Studies},
	author       = {Jansen, Anniek and Colombo, Sara},
	year         = {2022},
	booktitle    = {Extended Abstracts of the 2022 CHI Conference on Human Factors in Computing Systems},
	location     = {New Orleans, LA, USA},
	publisher    = {Association for Computing Machinery},
	address      = {New York, NY, USA},
	series       = {CHI EA '22},
	doi          = {10.1145/3491101.3519684},
	isbn         = {9781450391566},
	url          = {https://doi.org/10.1145/3491101.3519684},
	articleno    = {426},
	numpages     = {7}
}

@inproceedings{kapania_simulacrum_2025,
	title        = {Simulacrum of Stories: Examining Large Language Models as Qualitative Research Participants},
	author       = {Kapania, Shivani and Agnew, William and Eslami, Motahhare and Heidari, Hoda and Fox, Sarah E},
	year         = {2025},
	booktitle    = {Proceedings of the 2025 CHI Conference on Human Factors in Computing Systems},
	publisher    = {Association for Computing Machinery},
	address      = {New York, NY, USA},
	series       = {CHI '25},
	doi          = {10.1145/3706598.3713220},
	isbn         = {9798400713941},
	url          = {https://doi.org/10.1145/3706598.3713220},
	articleno    = {489},
	numpages     = {17}
}

@incollection{kjaerup_longitudinal_2021,
	title        = {Longitudinal studies in HCI research: a review of CHI publications from 1982--2019},
	author       = {Kj{\ae}rup, Maria and Skov, Mikael B and Nielsen, Peter Axel and Kjeldskov, Jesper and Gerken, Jens and Reiterer, Harald},
	year         = {2021},
	booktitle    = {Some Edited Volume Title or Proceedings (if applicable)},
	publisher    = {Springer},
	doi          = {10.1007/978-3-030-67322-2_2},
	url          = {https://doi.org/10.1007/978-3-030-67322-2_2}
}

@inproceedings{kjeldskov_was_2014,
	title        = {Was it worth the hassle? ten years of mobile HCI research discussions on lab and field evaluations},
	author       = {Kjeldskov, Jesper and Skov, Mikael B.},
	year         = {2014},
	booktitle    = {Proceedings of the 16th International Conference on Human-Computer Interaction with Mobile Devices \& Services},
	location     = {Toronto, ON, Canada},
	publisher    = {Association for Computing Machinery},
	address      = {New York, NY, USA},
	series       = {MobileHCI '14},
	pages        = {43–52},
	doi          = {10.1145/2628363.2628398},
	isbn         = {9781450330046},
	url          = {https://doi.org/10.1145/2628363.2628398},
	numpages     = {10}
}

@inproceedings{klemmer_suede_2000,
	title        = {Suede: a Wizard of Oz prototyping tool for speech user interfaces},
	author       = {Klemmer, Scott R. and Sinha, Anoop K. and Chen, Jack and Landay, James A. and Aboobaker, Nadeem and Wang, Annie},
	year         = {2000},
	booktitle    = {Proceedings of the 13th Annual ACM Symposium on User Interface Software and Technology},
	location     = {San Diego, California, USA},
	publisher    = {Association for Computing Machinery},
	address      = {New York, NY, USA},
	series       = {UIST '00},
	pages        = {1–10},
	doi          = {10.1145/354401.354406},
	isbn         = {1581132123},
	url          = {https://doi.org/10.1145/354401.354406},
	numpages     = {10}
}

@inproceedings{kolln_identifying_2022,
	title        = {Identifying User Experience Aspects for Voice User Interfaces with Intensive Users},
	author       = {K{\"o}lln, Kristina and Deutschl{\"a}nder, Jana and Klein, Andreas M and Rauschenberger, Maria and Winter, Dominique},
	year         = {2022},
	booktitle    = {WEBIST},
	pages        = {385--393},
	doi          = {10.5220/0011383300003318}
}

@article{kosch_risk_2024,
	title        = {Risk or Chance? Large Language Models and Reproducibility in HCI Research},
	author       = {Kosch, Thomas and Feger, Sebastian},
	year         = {2024},
	month        = oct,
	journal      = {Interactions},
	publisher    = {Association for Computing Machinery},
	address      = {New York, NY, USA},
	volume       = {31},
	number       = {6},
	pages        = {44–49},
	doi          = {10.1145/3695765},
	issn         = {1072-5520},
	url          = {https://doi.org/10.1145/3695765},
	issue_date   = {November - December 2024},
	numpages     = {6}
}

@article{krakowski_human-ai_2025,
	title        = {Human-AI agency in the age of generative AI},
	author       = {Krakowski, Sebastian},
	year         = {2025},
	journal      = {Information and Organization},
	publisher    = {Elsevier},
	volume       = {35},
	number       = {1},
	pages        = {100560},
	doi          = {10.1016/j.infoandorg.2025.100560}
}

@article{lam_empirical_2012,
	title        = {Empirical studies in information visualization: Seven scenarios},
	author       = {Lam, Heidi and Bertini, Enrico and Isenberg, Petra and Plaisant, Catherine and Carpendale, Sheelagh},
	year         = {2012},
	journal      = {IEEE transactions on visualization and computer graphics},
	publisher    = {IEEE},
	volume       = {18},
	number       = {9},
	pages        = {1520--1536},
	doi          = {10.1109/TVCG.2011.279}
}

@book{lazar_research_2010,
	title        = {Research {Methods} in {Human}-{Computer} {Interaction}},
	author       = {Lazar, Jonathan and Feng, Jinjuan Heidi and Hochheiser, Harry},
	year         = {2010},
	month        = jan,
	publisher    = {Wiley Publishing},
	isbn         = {978-0-470-72337-1}
}

@inproceedings{lee_storagechat_2025,
	title        = {StorageChat Timeline: A Generative AI-Based Art Appreciation System for Enhancing Immersion and Exploratory Experience},
	author       = {Lee, Chaeyeon and Lee, Chungnyeong and Kim, Sangyong and Choi, Yongsoon and Kim, Jusub},
	year         = {2025},
	booktitle    = {Proceedings of the Extended Abstracts of the CHI Conference on Human Factors in Computing Systems},
	location     = {},
	publisher    = {Association for Computing Machinery},
	address      = {New York, NY, USA},
	series       = {CHI EA '25},
	doi          = {10.1145/3706599.3721349},
	isbn         = {9798400713958},
	url          = {https://doi.org/10.1145/3706599.3721349},
	articleno    = {921},
	numpages     = {2}
}

@inproceedings{mackenzie_user_2015,
	title        = {User studies and usability evaluations: from research to products},
	shorttitle   = {User studies and usability evaluations},
	author       = {MacKenzie, I. Scott},
	year         = {2015},
	month        = jun,
	booktitle    = {Proceedings of the 41st {Graphics} {Interface} {Conference}},
	publisher    = {Canadian Information Processing Society},
	address      = {CAN},
	series       = {{GI} '15},
	pages        = {1--8},
	isbn         = {978-0-9947868-0-7},
	urldate      = {2025-06-04}
}

@incollection{mcgrath_methodology_1995,
	title        = {Methodology matters: Doing research in the behavioral and social sciences},
	author       = {McGrath, Joseph E},
	year         = {1995},
	booktitle    = {Readings in human--computer interaction},
	publisher    = {Elsevier},
	pages        = {152--169},
	doi          = {10.1016/B978-0-08-051574-8.50019-4},
	url          = {https://linkinghub.elsevier.com/retrieve/pii/B9780080515748500194}
}

@article{munzner_nested_2009,
	title        = {A nested model for visualization design and validation},
	author       = {Munzner, Tamara},
	year         = {2009},
	journal      = {IEEE transactions on visualization and computer graphics},
	publisher    = {IEEE},
	volume       = {15},
	number       = {6},
	pages        = {921--928},
	doi          = {10.1109/TVCG.2009.111},
	url          = {https://ieeexplore.ieee.org/document/5290695}
}

@article{murray-smith_what_2022,
	title        = {What simulation can do for HCI research},
	author       = {Murray-Smith, Roderick and Oulasvirta, Antti and Howes, Andrew and M\"{u}ller, J\"{o}rg and Ikkala, Aleksi and Bachinski, Miroslav and Fleig, Arthur and Fischer, Florian and Klar, Markus},
	year         = {2022},
	month        = nov,
	journal      = {Interactions},
	publisher    = {Association for Computing Machinery},
	address      = {New York, NY, USA},
	volume       = {29},
	number       = {6},
	pages        = {48–53},
	doi          = {10.1145/3564038},
	issn         = {1072-5520},
	url          = {https://doi.org/10.1145/3564038},
	issue_date   = {November - December 2022},
	numpages     = {6}
}

@inproceedings{park_we_2024,
	title        = {"We Are Visual Thinkers, Not Verbal Thinkers!": A Thematic Analysis of How Professional Designers Use Generative AI Image Generation Tools},
	author       = {Park, Hyerim and Eirich, Joscha and Luckow, Andre and Sedlmair, Michael},
	year         = {2024},
	booktitle    = {Proceedings of the 13th Nordic Conference on Human-Computer Interaction},
	location     = {Uppsala, Sweden},
	publisher    = {Association for Computing Machinery},
	address      = {New York, NY, USA},
	series       = {NordiCHI '24},
	doi          = {10.1145/3679318.3685370},
	isbn         = {9798400709661},
	url          = {https://doi.org/10.1145/3679318.3685370},
	articleno    = {35},
	numpages     = {14}
}

@article{petzoldt_critical_2014,
	title        = {The critical tracking task: a potentially useful method to assess driver distraction?},
	author       = {Petzoldt, Tibor and Bellem, Hanna and Krems, Josef F},
	year         = {2014},
	journal      = {Human factors},
	publisher    = {Sage Publications Sage CA: Los Angeles, CA},
	volume       = {56},
	number       = {4},
	pages        = {789--808},
	doi          = {10.1177/0018720813501864},
	url          = {https://journals.sagepub.com/doi/10.1177/0018720813501864}
}

@inproceedings{plaisant_challenge_2004,
	title        = {The challenge of information visualization evaluation},
	author       = {Plaisant, Catherine},
	year         = {2004},
	booktitle    = {Proceedings of the Working Conference on Advanced Visual Interfaces},
	location     = {Gallipoli, Italy},
	publisher    = {Association for Computing Machinery},
	address      = {New York, NY, USA},
	series       = {AVI '04},
	pages        = {109–116},
	doi          = {10.1145/989863.989880},
	isbn         = {1581138679},
	url          = {https://doi.org/10.1145/989863.989880},
	numpages     = {8}
}

@inproceedings{poppe_evaluating_2007,
	title        = {Evaluating the future of HCI: challenges for the evaluation of emerging applications},
	author       = {Poppe, Ronald and Rienks, Rutger and van Dijk, Betsy},
	year         = {2007},
	booktitle    = {Artificial Intelligence for Human Computing: ICMI 2006 and IJCAI 2007 International Workshops, Banff, Canada, November 3, 2006, Hyderabad, India, January 6, 2007, Revised Selected and Invited Papers},
	pages        = {234--250},
	doi          = {10.1007/978-3-540-72348-6_12},
	organization = {Springer}
}

@inproceedings{pourasad_does_2025,
	title        = {Does GenAI Make Usability Testing Obsolete?},
	author       = {Pourasad, Ali Ebrahimi and Maalej, Walid},
	year         = {2025},
	booktitle    = {2025 IEEE/ACM 47th International Conference on Software Engineering (ICSE)},
	publisher    = {IEEE},
	pages        = {437--449},
	doi          = {10.1109/ICSE55347.2025.00138},
	url          = {https://ieeexplore.ieee.org/document/11029918}
}

@book{preece_human-computer_1994,
	title        = {Human-{Computer} {Interaction}},
	author       = {Preece, Jenny and Rogers, Yvonne and Sharp, Helen and Benyon, David and Holland, Simon and Carey, Tom},
	year         = {1994},
	month        = mar,
	publisher    = {Addison-Wesley Longman Ltd.},
	address      = {GBR},
	isbn         = {978-0-201-62769-5}
}

@article{ravera_usability_2025,
	title        = {On the usability of generative AI: Human generative AI},
	author       = {Ravera, Anna and Gena, Cristina},
	year         = {2025},
	journal      = {arXiv preprint arXiv:2502.17714},
	doi          = {10.48550/arXiv.2502.17714},
	url          = {https://arxiv.org/abs/2502.17714},
	urldate      = {2025-06-04}
}

@inproceedings{rayan_exploring_2024,
	title        = {Exploring the Potential for Generative AI-based Conversational Cues for Real-Time Collaborative Ideation},
	author       = {Rayan, Jude and Kanetkar, Dhruv and Gong, Yifan and Yang, Yuewen and Palani, Srishti and Xia, Haijun and Dow, Steven P.},
	year         = {2024},
	booktitle    = {Proceedings of the 16th Conference on Creativity \& Cognition},
	location     = {Chicago, IL, USA},
	publisher    = {Association for Computing Machinery},
	address      = {New York, NY, USA},
	series       = {C\&C '24},
	pages        = {117–131},
	doi          = {10.1145/3635636.3656184},
	isbn         = {9798400704857},
	url          = {https://doi.org/10.1145/3635636.3656184},
	numpages     = {15}
}

@article{riek_wizard_2012,
	title        = {Wizard of {Oz} studies in {HRI}: a systematic review and new reporting guidelines},
	shorttitle   = {Wizard of {Oz} studies in {HRI}},
	author       = {Riek, Laurel D.},
	year         = {2012},
	month        = jul,
	journal      = {J. Hum.-Robot Interact.},
	volume       = {1},
	number       = {1},
	pages        = {119--136},
	doi          = {10.5898/JHRI.1.1.Riek},
	url          = {https://dl.acm.org/doi/10.5898/JHRI.1.1.Riek},
	urldate      = {2025-12-21}
}

@article{rind_whys_2011,
	title        = {Some {Whys} and {Hows} of {Experiments} in {Human}–{Computer} {Interaction}},
	author       = {Rind, Alexander},
	year         = {2011},
	month        = jan,
	journal      = {Foundations and Trends® in Human-Computer Interaction},
	volume       = {5},
	pages        = {299--373},
	doi          = {10.1561/1100000043}
}

@incollection{rogers_case_2017,
	title        = {Case {Studies}: {Designing} and {Evaluating} {Technologies} for {Use} in the {Wild}},
	shorttitle   = {Case {Studies}},
	author       = {Rogers, Yvonne and Marshall, Paul},
	year         = {2017},
	booktitle    = {Research in the {Wild}},
	publisher    = {Springer International Publishing},
	address      = {Cham},
	pages        = {33--67},
	doi          = {10.1007/978-3-031-02220-3_4},
	isbn         = {978-3-031-02220-3},
	url          = {https://doi.org/10.1007/978-3-031-02220-3_4},
	urldate      = {2025-06-30},
	language     = {en},
	editor       = {Rogers, Yvonne and Marshall, Paul}
}

@book{rogers_research_2017,
	title        = {Research in the {Wild}},
	author       = {Rogers, Yvonne and Marshall, Paul and Carroll, John M.},
	year         = {2017},
	month        = mar,
	publisher    = {Morgan \& Claypool Publishers},
	isbn         = {978-1-62705-692-2}
}

@article{rong_towards_2024,
	title        = {Towards {Human}-{Centered} {Explainable} {AI}: {A} {Survey} of {User} {Studies} for {Model} {Explanations}},
	shorttitle   = {Towards {Human}-{Centered} {Explainable} {AI}},
	author       = {Rong, Yao and Leemann, Tobias and Nguyen, Thai-Trang and Fiedler, Lisa and Qian, Peizhu and Unhelkar, Vaibhav and Seidel, Tina and Kasneci, Gjergji and Kasneci, Enkelejda},
	year         = {2024},
	month        = apr,
	journal      = {IEEE Transactions on Pattern Analysis and Machine Intelligence},
	volume       = {46},
	number       = {4},
	pages        = {2104--2122},
	doi          = {10.1109/TPAMI.2023.3331846},
	issn         = {1939-3539},
	url          = {https://ieeexplore.ieee.org/document/10316181/},
	urldate      = {2025-06-30}
}

@inproceedings{schmidt_evaluation_2021,
	title        = {Evaluation in Human-Computer Interaction – Beyond Lab Studies},
	author       = {Schmidt, Albrecht and Alt, Florian and M\"{a}kel\"{a}, Ville},
	year         = {2021},
	booktitle    = {Extended Abstracts of the 2021 CHI Conference on Human Factors in Computing Systems},
	location     = {Yokohama, Japan},
	publisher    = {Association for Computing Machinery},
	address      = {New York, NY, USA},
	series       = {CHI EA '21},
	doi          = {10.1145/3411763.3445022},
	isbn         = {9781450380959},
	url          = {https://doi.org/10.1145/3411763.3445022},
	articleno    = {142},
	numpages     = {4}
}

@article{sedlmair_design_2012,
	title        = {Design study methodology: Reflections from the trenches and the stacks},
	author       = {Sedlmair, Michael and Meyer, Miriah and Munzner, Tamara},
	year         = {2012},
	journal      = {IEEE transactions on visualization and computer graphics},
	publisher    = {IEEE},
	volume       = {18},
	number       = {12},
	pages        = {2431--2440},
	doi          = {10.1109/TVCG.2012.213},
	url          = {http://ieeexplore.ieee.org/document/6327248/}
}

@article{shi_hci-centric_2024,
	title        = {An HCI-centric survey and taxonomy of human-generative-AI interactions},
	author       = {Shi, Jingyu and Jain, Rahul and Doh, Hyungjun and Suzuki, Ryo and Ramani, Karthik},
	year         = {2023},
	journal      = {arXiv preprint arXiv:2310.07127},
	doi          = {10.48550/arXiv.2310.07127},
	url          = {http://arxiv.org/abs/2310.07127}
}

@inproceedings{silkhi_comparative_2024,
	title        = {Comparative Analysis of Rule-Based Chatbot Development Tools for Education Orientation: A RAD Approach},
	author       = {Silkhi, Hassan and Bakkas, Brahim and Housni, Khalid},
	year         = {2024},
	booktitle    = {Proceedings of the 7th International Conference on Networking, Intelligent Systems and Security},
	location     = {Meknes, AA, Morocco},
	publisher    = {Association for Computing Machinery},
	address      = {New York, NY, USA},
	series       = {NISS '24},
	doi          = {10.1145/3659677.3659825},
	isbn         = {9798400709296},
	url          = {https://doi.org/10.1145/3659677.3659825},
	articleno    = {51},
	numpages     = {7}
}

@article{simkute_ironies_2025,
	title        = {Ironies of generative AI: understanding and mitigating productivity loss in Human-AI interaction},
	author       = {Simkute, Auste and Tankelevitch, Lev and Kewenig, Viktor and Scott, Ava Elizabeth and Sellen, Abigail and Rintel, Sean},
	year         = {2025},
	journal      = {International Journal of Human--Computer Interaction},
	publisher    = {Taylor \& Francis},
	volume       = {41},
	number       = {5},
	pages        = {2898--2919},
	doi          = {10.1080/10447318.2024.2405782},
	url          = {https://doi.org/10.1080/10447318.2024.2405782}
}

@article{simonen_exploration_2025,
	title        = {An Initial Exploration of Default Images in Text-to-Image Generation},
	author       = {Simonen, Hannu and Kiviniemi, Atte and Oppenlaender, Jonas},
	year         = {2025},
	journal      = {arXiv preprint arXiv:2505.09166},
	doi          = {10.48550/arXiv.2505.09166},
	url          = {http://arxiv.org/abs/2505.09166}
}

@article{singh_survey_2025,
	title        = {A survey on chatbots and large language models: Testing and evaluation techniques},
	author       = {Singh, Sonali Uttam and Namin, Akbar Siami},
	year         = {2025},
	journal      = {Natural Language Processing Journal},
	publisher    = {Elsevier},
	pages        = {100128},
	doi          = {10.1016/j.nlp.2025.100128},
	url          = {https://www.sciencedirect.com/science/article/pii/S2949719125000044}
}

@inproceedings{steinfeld_oz_2009,
	title        = {The oz of wizard: simulating the human for interaction research},
	author       = {Steinfeld, Aaron and Jenkins, Odest Chadwicke and Scassellati, Brian},
	year         = {2009},
	booktitle    = {Proceedings of the 4th ACM/IEEE International Conference on Human Robot Interaction},
	location     = {La Jolla, California, USA},
	publisher    = {Association for Computing Machinery},
	address      = {New York, NY, USA},
	series       = {HRI '09},
	pages        = {101–108},
	doi          = {10.1145/1514095.1514115},
	isbn         = {9781605584041},
	url          = {https://doi.org/10.1145/1514095.1514115},
	numpages     = {8}
}

@inproceedings{stigall_systematic_2020,
	title        = {A systematic review of human factors literature about voice user interfaces and older adults},
	author       = {Stigall, Brodrick and Caine, Kelly},
	year         = {2020},
	booktitle    = {Proceedings of the Human Factors and Ergonomics Society Annual Meeting},
	volume       = {64},
	number       = {1},
	pages        = {13--17},
	doi          = {10.1177/1071181320641004},
	url          = {https://journals.sagepub.com/action/showAbstract},
	organization = {SAGE Publications Sage CA: Los Angeles, CA}
}

@article{storms_transparency_2022,
	title        = {'Transparency is Meant for Control' and Vice Versa: Learning from Co-designing and Evaluating Algorithmic News Recommenders},
	author       = {Storms, Elias and Alvarado, Oscar and Monteiro-Krebs, Luciana},
	year         = {2022},
	month        = nov,
	journal      = {Proc. ACM Hum.-Comput. Interact.},
	publisher    = {Association for Computing Machinery},
	address      = {New York, NY, USA},
	volume       = {6},
	number       = {CSCW2},
	doi          = {10.1145/3555130},
	url          = {https://doi.org/10.1145/3555130},
	issue_date   = {November 2022},
	articleno    = {405},
	numpages     = {24}
}

@inproceedings{su_enhancing_2024,
	title        = {Enhancing User Experience Evaluation of Graphic Art Style Games through Collaboration with Generative AI},
	author       = {Su, Jiajia and He, Zhongjun},
	year         = {2024},
	booktitle    = {Proceedings of the 2024 5th International Conference on Computer Science and Management Technology},
	location     = {},
	publisher    = {Association for Computing Machinery},
	address      = {New York, NY, USA},
	series       = {ICCSMT '24},
	pages        = {31–38},
	doi          = {10.1145/3708036.3708042},
	isbn         = {9798400709999},
	url          = {https://doi.org/10.1145/3708036.3708042},
	numpages     = {8}
}

@inproceedings{tan_rule-based_2025,
	title        = {Rule-Based vs. AI-Driven: Comparing PolyAQG Framework and Generative AI Models},
	author       = {Tan, Tee Hean},
	year         = {2025},
	booktitle    = {Proceedings of the 2024 8th International Conference on Natural Language Processing and Information Retrieval},
	location     = {},
	publisher    = {Association for Computing Machinery},
	address      = {New York, NY, USA},
	series       = {NLPIR '24},
	pages        = {298–303},
	doi          = {10.1145/3711542.3711583},
	isbn         = {9798400717383},
	url          = {https://doi.org/10.1145/3711542.3711583},
	numpages     = {6}
}

@inproceedings{todi_adapting_2021,
	title        = {Adapting User Interfaces with Model-based Reinforcement Learning},
	author       = {Todi, Kashyap and Bailly, Gilles and Leiva, Luis and Oulasvirta, Antti},
	year         = {2021},
	booktitle    = {Proceedings of the 2021 CHI Conference on Human Factors in Computing Systems},
	location     = {Yokohama, Japan},
	publisher    = {Association for Computing Machinery},
	address      = {New York, NY, USA},
	series       = {CHI '21},
	doi          = {10.1145/3411764.3445497},
	isbn         = {9781450380966},
	url          = {https://doi.org/10.1145/3411764.3445497},
	articleno    = {573},
	numpages     = {13}
}

@inproceedings{van_turnhout_design_2014,
	title        = {Design patterns for mixed-method research in HCI},
	author       = {van Turnhout, Koen and Bennis, Arthur and Craenmehr, Sabine and Holwerda, Robert and Jacobs, Marjolein and Niels, Ralph and Zaad, Lambert and Hoppenbrouwers, Stijn and Lenior, Dick and Bakker, Ren\'{e}},
	year         = {2014},
	booktitle    = {Proceedings of the 8th Nordic Conference on Human-Computer Interaction: Fun, Fast, Foundational},
	location     = {Helsinki, Finland},
	publisher    = {Association for Computing Machinery},
	address      = {New York, NY, USA},
	series       = {NordiCHI '14},
	pages        = {361–370},
	doi          = {10.1145/2639189.2639220},
	isbn         = {9781450325424},
	url          = {https://doi.org/10.1145/2639189.2639220},
	numpages     = {10}
}

@inproceedings{volkel_what_2020,
	title        = {What is "intelligent" in intelligent user interfaces? a meta-analysis of 25 years of IUI},
	author       = {V\"{o}lkel, Sarah Theres and Schneegass, Christina and Eiband, Malin and Buschek, Daniel},
	year         = {2020},
	booktitle    = {Proceedings of the 25th International Conference on Intelligent User Interfaces},
	location     = {Cagliari, Italy},
	publisher    = {Association for Computing Machinery},
	address      = {New York, NY, USA},
	series       = {IUI '20},
	pages        = {477–487},
	doi          = {10.1145/3377325.3377500},
	isbn         = {9781450371186},
	url          = {https://doi.org/10.1145/3377325.3377500},
	numpages     = {11}
}

@inproceedings{wolf_role_1989,
	title        = {The role of laboratory  experiments in HCI: help, hindrance, or ho-hum?},
	author       = {Wolf, C. G. and Carroll, J. M. and Landauer, T. K. and John, B. E. and Whiteside, J.},
	year         = {1989},
	booktitle    = {Proceedings of the SIGCHI Conference on Human Factors in Computing Systems},
	publisher    = {Association for Computing Machinery},
	address      = {New York, NY, USA},
	series       = {CHI '89},
	pages        = {265–268},
	doi          = {10.1145/67449.67500},
	isbn         = {0897913019},
	url          = {https://doi.org/10.1145/67449.67500},
	numpages     = {4}
}

@article{woolrych_ingredients_2011,
	title        = {Ingredients and meals rather than recipes: A proposal for research that does not treat usability evaluation methods as indivisible wholes},
	author       = {Woolrych, Alan and Hornb{\ae}k, Kasper and Fr{\o}kj{\ae}r, Erik and Cockton, Gilbert},
	year         = {2011},
	journal      = {International Journal of Human-Computer Interaction},
	publisher    = {Taylor \& Francis},
	volume       = {27},
	number       = {10},
	pages        = {940--970},
	doi          = {10.1080/10447318.2011.555314},
	url          = {http://www.tandfonline.com/doi/abs/10.1080/10447318.2011.555314}
}

@inproceedings{yan_human-ai_2024,
	title        = {Human-AI Collaboration in Thematic Analysis using ChatGPT: A User Study and Design Recommendations},
	author       = {Yan, Lixiang and Echeverria, Vanessa and Fernandez-Nieto, Gloria Milena and Jin, Yueqiao and Swiecki, Zachari and Zhao, Linxuan and Ga\v{s}evi\'{c}, Dragan and Martinez-Maldonado, Roberto},
	year         = {2024},
	booktitle    = {Extended Abstracts of the CHI Conference on Human Factors in Computing Systems},
	location     = {Honolulu, HI, USA},
	publisher    = {Association for Computing Machinery},
	address      = {New York, NY, USA},
	series       = {CHI EA '24},
	doi          = {10.1145/3613905.3650732},
	isbn         = {9798400703317},
	url          = {https://doi.org/10.1145/3613905.3650732},
	articleno    = {191},
	numpages     = {7}
}

@inproceedings{yin_understanding_2019,
	title        = {Understanding the Effect of Accuracy on Trust in Machine Learning Models},
	author       = {Yin, Ming and Wortman Vaughan, Jennifer and Wallach, Hanna},
	year         = {2019},
	booktitle    = {Proceedings of the 2019 CHI Conference on Human Factors in Computing Systems},
	location     = {Glasgow, Scotland Uk},
	publisher    = {Association for Computing Machinery},
	address      = {New York, NY, USA},
	series       = {CHI '19},
	pages        = {1–12},
	doi          = {10.1145/3290605.3300509},
	isbn         = {9781450359702},
	url          = {https://doi.org/10.1145/3290605.3300509},
	numpages     = {12}
}

@inproceedings{yun_generative_2025,
	title        = {Generative AI in Knowledge Work: Design Implications for Data Navigation and Decision-Making},
	author       = {Yun, Bhada and Feng, Dana and Chen, Ace S. and Nikzad, Afshin and Salehi, Niloufar},
	year         = {2025},
	booktitle    = {Proceedings of the 2025 CHI Conference on Human Factors in Computing Systems},
	location     = {},
	publisher    = {Association for Computing Machinery},
	address      = {New York, NY, USA},
	series       = {CHI '25},
	doi          = {10.1145/3706598.3713337},
	isbn         = {9798400713941},
	url          = {https://doi.org/10.1145/3706598.3713337},
	articleno    = {634},
	numpages     = {19}
}

@inproceedings{yun_user_2025,
	title        = {User Experience with LLM-powered Conversational Recommendation Systems: A Case of Music Recommendation},
	author       = {Yun, Sojeong and Lim, Youn-kyung},
	year         = {2025},
	booktitle    = {Proceedings of the 2025 CHI Conference on Human Factors in Computing Systems},
	location     = {},
	publisher    = {Association for Computing Machinery},
	address      = {New York, NY, USA},
	series       = {CHI '25},
	doi          = {10.1145/3706598.3713347},
	isbn         = {9798400713941},
	url          = {https://doi.org/10.1145/3706598.3713347},
	articleno    = {898},
	numpages     = {15}
}

@inproceedings{zamfirescu-pereira_why_2023,
	title        = {Why Johnny Can’t Prompt: How Non-AI Experts Try (and Fail) to Design LLM Prompts},
	author       = {Zamfirescu-Pereira, J.D. and Wong, Richmond Y. and Hartmann, Bjoern and Yang, Qian},
	year         = {2023},
	booktitle    = {Proceedings of the 2023 CHI Conference on Human Factors in Computing Systems},
	location     = {Hamburg, Germany},
	publisher    = {Association for Computing Machinery},
	address      = {New York, NY, USA},
	series       = {CHI '23},
	doi          = {10.1145/3544548.3581388},
	isbn         = {9781450394215},
	url          = {https://doi.org/10.1145/3544548.3581388},
	articleno    = {437},
	numpages     = {21}
}

@inproceedings{zheng_evalignux_2025,
	title        = {EvAlignUX: Advancing UX Evaluation through LLM-Supported Metrics Exploration},
	author       = {Zheng, Qingxiao and Chen, Minrui and Sharma, Pranav and Tang, Yiliu and Oswal, Mehul and Liu, Yiren and Huang, Yun},
	year         = {2025},
	booktitle    = {Proceedings of the 2025 CHI Conference on Human Factors in Computing Systems},
	location     = {},
	publisher    = {Association for Computing Machinery},
	address      = {New York, NY, USA},
	series       = {CHI '25},
	doi          = {10.1145/3706598.3714045},
	isbn         = {9798400713941},
	url          = {https://doi.org/10.1145/3706598.3714045},
	articleno    = {1051},
	numpages     = {25}
}

\appendix

\section{Appendix}
\subsection{Initial Stage of Analysis}
\label{app:initial_analysis}
\begin{figure*}[h]
    \centering
    \includegraphics[width=\linewidth]{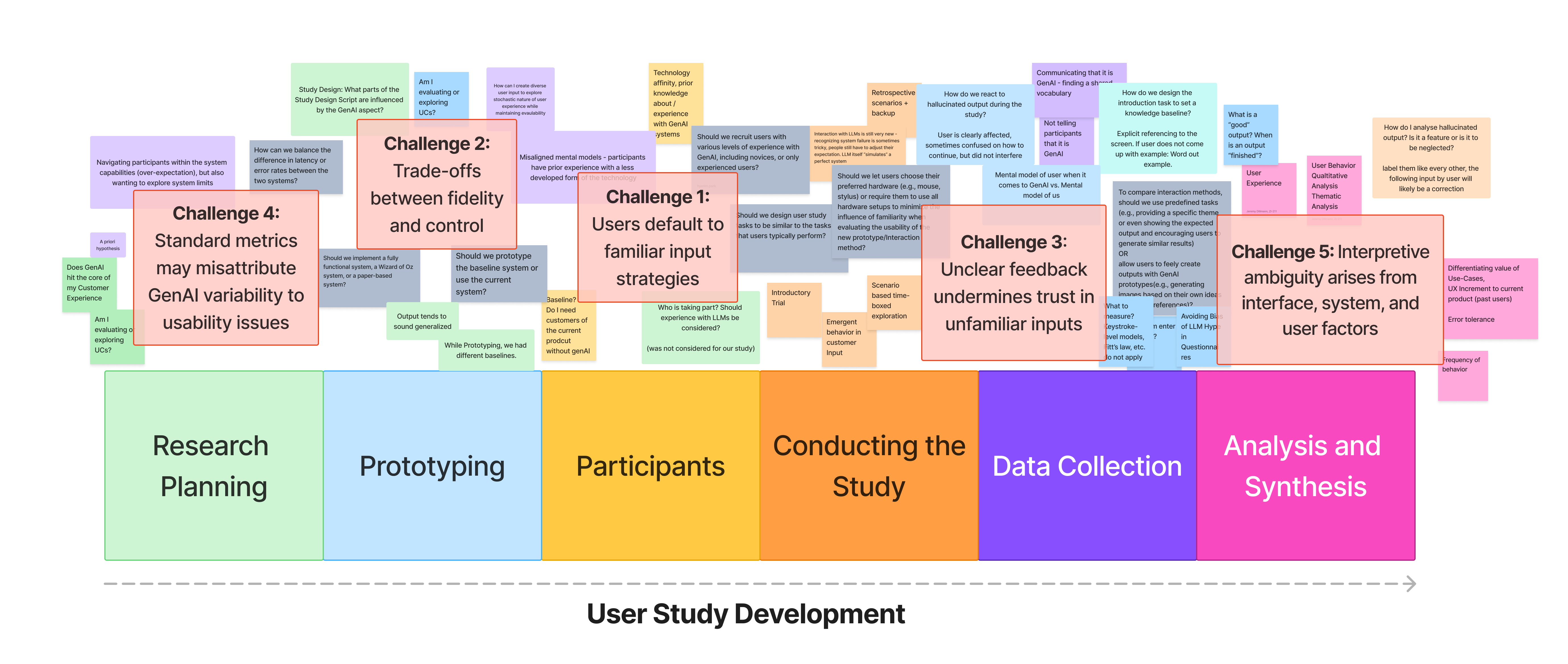}
    \caption{Overview of the initial collaborative workspace used during an early stage of analysis. Methodological observations were externalized and provisionally grouped to support shared sensemaking and explore emerging patterns. These materials were iteratively added, merged, and reorganized to explore potential subthemes and overarching patterns, informed by principles of thematic analysis~\cite{braun_using_2006}. Challenge labels (C1--C5) were added to make the connection between early affinity notes and the resulting challenges explicit.}
    \label{fig:initial_analysis}
    \Description{The figure shows a collaborative workspace used during an early stage of qualitative analysis. A large canvas contains clusters of color-coded notes representing methodological observations and affinity notes. These clusters are arranged along a horizontal timeline of user study development phases---research planning, prototyping, participants, procedure, data collection, and analysis and synthesis. Higher-level labels mark five emerging methodological challenges (C1--C5), positioned above related clusters to show how early observations were grouped and iteratively reorganized into candidate themes. The figure illustrates the transition from granular observations to provisional challenge categories during sensemaking.}
\end{figure*}

The initial collaborative workspace was used in the first round of analysis. In this phase, we brainstormed and collected methodological observations from four user studies, organizing them according to study phases (e.g., research planning, prototyping, data collection, and analysis). This initial mapping served as an analytic scaffold, helping us identify where and when methodological issues emerged and supporting early sensemaking around recurring patterns (see \autoref{fig:initial_analysis}). 

\subsection{Overview of Case Studies}
\label{app:cases}
This appendix provides an overview of the four lab-based user studies analyzed in our multi-case reflection. Each case represents a distinct evaluation context for GenAI systems, ranging from early concept exploration to studies involving fully functional prototypes. The studies vary in domain, prototype fidelity, and participant group, collectively illustrating recurring methodological challenges encountered when evaluating GenAI systems in controlled lab settings.

\subsubsection{Case Study A: An LLM-Based Conversational Car Assistant}
\label{sec:case_a}
\paragraph{Study Objective}
This study investigated how users interact with an LLM-based conversational car assistant during various driving-related tasks. We examined conversation flow (single-turn versus multi-turn), language style (command-based versus natural language), task completion, and recovery from system errors. Additionally, we assessed distraction levels during interaction and overall usability. We were particularly interested in how users responded when the system could not handle certain requests that were not yet implemented in the prototype.

\paragraph{Study Procedure}
We recruited 30 participants for our study, which was conducted in a standing vehicle with participants seated in the driver's seat. Participants were given structured tasks related to driving and car controls, each followed by a post-task interview and evaluation. The tasks included navigating to a destination with a stop along the route, controlling the windows and lights via speech, asking about car functionalities typically covered in the car manual, and engaging in free conversation on a topic of their choice. These tasks were introduced in a way that guided participants while still allowing them as much freedom as possible, including the option to go beyond the system's limits. Additionally, participants completed the Critical Tracking Task~\cite{petzoldt_critical_2014} on a screen positioned in front of the car to simulate driving-equivalent cognitive load for half of the task duration.

\paragraph{Data Collection and Analysis}
We collected qualitative data through post-task interviews, as well as interviews conducted at the beginning and end of each session. Quantitative measures included the deviation in the Critical Tracking Task, feedback gathered using Likert-scale items from the UEQ, and system errors recorded through observations made by the study team. Moreover, we logged user utterances, system replies, and the number of conversational turns.

\subsubsection{Case Study B: An LLM-Based Conversational Navigation Assistant Referencing the Display}
\paragraph{Study Objective}
\label{sec:case_b}
In this case, we investigated user interaction patterns in a multimodal LLM-based VUI within the automotive navigation context. We used a system that integrates GPT-4's multimodal capabilities with screenshots of the car's central display. Participants could ask questions about visual elements on the map or other GUI components shown on the central display. Using this system in a stationary vehicle, we conducted a user study with 21 participants. Through structured tasks and post-task questionnaires, we analyzed how users verbally described visual elements, including map features and interface icons, and assessed system usability using the SUS. We created a taxonomy that categorizes the linguistic structures participants used when referencing visual elements through the VUI.

\paragraph{Study Procedure}
All participants interacted with the multimodal LLM-based system via speech to complete three tasks while seated in the driver's seat of a stationary vehicle. The three tasks varied in goals and interaction complexity. The first involved searching for charging stations within the navigation system. The second required participants to reference a lake shown on the map. The final task was an open exploration activity, where participants could ask questions about the elements displayed on the interface. This procedure allowed us to examine how users structured spatial references across different interaction types.

\paragraph{Data Collection and Analysis}
We collected transcripts of user--LLM interactions along with the contextual images provided to the model. Post-task interviews yielded both quantitative usability assessments via the SUS questionnaire and qualitative insights through semi-structured discussions. To analyze user behavior, we identified all utterances containing spatial references and applied thematic analysis~\cite{blandford_qualitative_2016}, resulting in a taxonomy of reference types and interaction patterns.

\paragraph{Key Differences from Case Study A}
This case builds on the conversational in-car assistant studied in Case Study A by focusing on how users reference visual elements in the GUI. This focus allowed us to examine users' referencing strategies and how successful references affect usability. By narrowing the interaction context, this study highlights challenges that arise specifically when conversational input is grounded in shared visual representations.

\subsubsection{Case Study C: A Paper-Based Exploration of Visual Input Methods for GenAI Image Tools}
\label{sec:case_c}

\paragraph{Study Objectives}
This study explored alternative interaction methods for GenAI image tools, which are typically centered on text prompts. To support more visual forms of input, we introduced scribble- and annotation-based interaction techniques that allow users to draw, handwrite, or annotate visual elements directly onto images, aligning with common design practices.

\paragraph{Prototype Design and Development}
We designed a hybrid input interface combining text prompts with scribbles and annotations. The front end was implemented using JavaScript and React and connected to several GenAI backends (Stable Diffusion, DALL·E~2, and GPT-4o). Early testing revealed latency and recognition errors in freehand input, which hindered the isolation of interaction behavior from system performance. We therefore adopted a paper-based prototype to simulate these interactions, enabling controlled evaluation of input strategies without interference from model variability.

\paragraph{Study Procedure}
A qualitative study was conducted with seven professional designers. Each participant compared three input methods (text prompts, scribbles, and annotations) across six design tasks. Five tasks focused on predefined refinement categories (adding objects, increasing complexity, making global changes, adjusting layout, and modifying texture), and one was open-ended. Participants used pen or keyboard input freely. A think-aloud protocol and post-task interviews were used to capture reasoning and reflections.

\paragraph{Data Collection and Analysis}
Qualitative data were collected from three sources: think-aloud protocols, post-task interviews, and user-generated artifacts (e.g., annotated sketches and text prompts). Think-aloud sessions captured participants' real-time prompting strategies, while interviews provided retrospective reflections on usability, preferences, and the perceived value of each input method. No structured questionnaires or quantitative metrics (e.g., Likert scales or task completion times) were used. Instead, the analysis was grounded in inductive thematic analysis~\cite{braun_using_2006}, focusing on patterns in behavior, input preferences, and recurring challenges across participants.

\subsubsection{Case Study D: A Functional Prototype User Study for a GenAI Image Generation Tool}
\label{sec:case_d}
\paragraph{Study Objectives}
Building on the findings from the earlier paper-based study (Case Study C), this follow-up study evaluated user interactions with a fully functional prototype of the GenAI image tool. While the previous study focused on input preferences in a static, controlled setup, this study observed how designers interacted with the system in real-time. The goal was to understand how real-time generative feedback affected user strategies, tool preferences, and iterative design behavior, as well as whether previously reported preferences for visual input methods persisted under dynamic feedback conditions.

\paragraph{Study Procedure}
This study expanded on the earlier paper-based work by introducing an interactive system that generates images in real time in response to user input. While Case Study C employed a low-fidelity prototype to investigate preferences in a fully controlled setting, Case Study D introduced a fully functional system with real-time, generative output, which enabled observation of how designers adapted their strategies dynamically to the system's stochastic behavior. Latency, error handling, and responsiveness---factors abstracted away in the previous study---became key aspects of this evaluation.

\paragraph{Data Collection and Analysis}
Participants included professional designers and design students who performed image refinement and creation tasks using text, scribbles, and annotations as input modalities. We collected usage logs, screen recordings, and post-task interview data. Quantitative measures included input completion times, the NASA-TLX, the UEQ, and a custom survey assessing perceived intent alignment. Quantitative results were used descriptively to contextualize the findings, while qualitative data were analyzed to capture behavioral adaptation and iterative strategies during live system interaction.

\paragraph{Key Differences from Case Study C}
Compared to Case Study C, this study examined real-time interaction with a fully functional GenAI system rather than simulated input. This enabled observation of how live feedback shaped iteration, trust, and engagement, offering insights into the usability and methodological implications of evaluating functional GenAI image tools in controlled settings.

\end{document}